\documentclass[10pt,journal,compsoc]{IEEEtran}

\usepackage{cite}
\usepackage{graphicx}
\usepackage{subcaption}
\usepackage{epsfig}
\usepackage{epstopdf}
\usepackage{color}
\usepackage{multirow}
\usepackage{booktabs}
\usepackage{amsmath,amsthm}
\usepackage{rotating}
\usepackage{algorithm}
\usepackage{algorithmic}
\usepackage{array}
\usepackage{listings}
\usepackage{animate}
\definecolor{dkgreen}{rgb}{0,0.6,0}
\definecolor{gray}{rgb}{0.5,0.5,0.5}
\definecolor{mauve}{rgb}{0.58,0,0.82}

\ifCLASSINFOpdf
\else
\fi
\begin{document}
\title{Echo: An Edge-Centric Code Offloading System with Quality of Service Guarantee}

\author{Li~Lin,

         Xiaofei~Liao,~\IEEEmembership{Member, ~IEEE,}

\IEEEcompsocitemizethanks{\IEEEcompsocthanksitem Li~Lin,~Xiaofei~Liao*(Corresponding author), are with the Service Computing Technology and System Lab, Cluster and Grid Computing Lab, and School of Computer Science and Technology, Huazhong University of Science and Technology, Wuhan 430074, China.}

\thanks}

\IEEEtitleabstractindextext{%
\begin{abstract}
Code offloading is promising to accelerate mobile applications and save energy of mobile devices by shifting some computation to cloud. However, existing code offloading systems suffer from a long communication delay between mobile devices and cloud. To address this challenge, in this paper, we consider to deploy edge nodes in the proximity of mobile devices, and study how they benefit code offloading. We design an edge-centric code offloading system, called Echo, over a three-layer computing hierarchy consisting of mobile devices, edge and cloud. A critical problem needs to be addressed by Echo is to decide which method should be offloaded to which computing platform (edge or cloud). Different from existing offloading systems that let mobile devices individually make offloading decisions, Echo implements a centralized decision engine at the edge node. This edge-centric design can fully exploit the limited hardware resources at the edge to provide an offloading service with Quality of Service guarantee. Furthermore, we propose some novel mechanisms, e.g., lazy object transmission and differential object update, to further improve system performance. The results of a small-scale real deployment and trace-driven simulations show that Echo significantly outperforms existing code offloading systems at both execution time and energy consumption.
\end{abstract}

}

\maketitle

\IEEEpeerreviewmaketitle

\section{Introduction}
\label{sec_introduction}
Mobile devices have evolved significantly with faster CPU and larger memory in recent years. However, they are still constrained to run large applications, such as augmented reality and games, due to limited hardware resources and battery capacity. Code offloading~\cite{MAUI,CloneCloud,ThinkAir} has been proposed to conquer this challenge by shifting some computation tasks of mobile devices to cloud. It can potentially save battery energy of mobile devices and accelerate mobile applications by using powerful hardware at the cloud. However, cloud is usually located geographically far from mobile devices, leading to limited network bandwidth and long transmission latency, which become the main obstacles to wide-scale deployment of code offloading.

Edge computing~\cite{anysee, EdgeComputing} emerges as a promising paradigm that deploys a number of modest-size computing nodes in the proximity of mobile devices, so that the computing requests of mobile devices can be quickly served with low latency. Many research efforts have been made to exploit the benefits of edge computing, such as the deep neural networks over end devices, the edge and the cloud~\cite{DNNOverEdge}, multi-level IoT systems~\cite{IoTOverEdge}, vehicular computing~\cite{VehicularOverEdge}, online algorithms for service reconfiguration at edge~\cite{HOU2016}, and the deployment of edge nodes~\cite{Tong_2016}. However, the integration of code offloading and edge computing with the performance guarantee is still an open challenge.

%Teerapittayanon et al. ~\cite{DNNOverEdge} have proposed a scalable distributed computing architecture of deep neural networks over end devices, the edge and the cloud. Nastic et al.~\cite{IoTOverEdge} have built a scalable multi-level provisioning of IoT systems, including the edge and the cloud. Hou et al. \cite{HOU2016} have designed online algorithms for service reconfiguration at edge. Tong et al. \cite{Tong_2016} have proposed a tree hierarchy for edge deployment.

\begin{figure}[tb]
\centering
\includegraphics[width=0.35\textwidth]{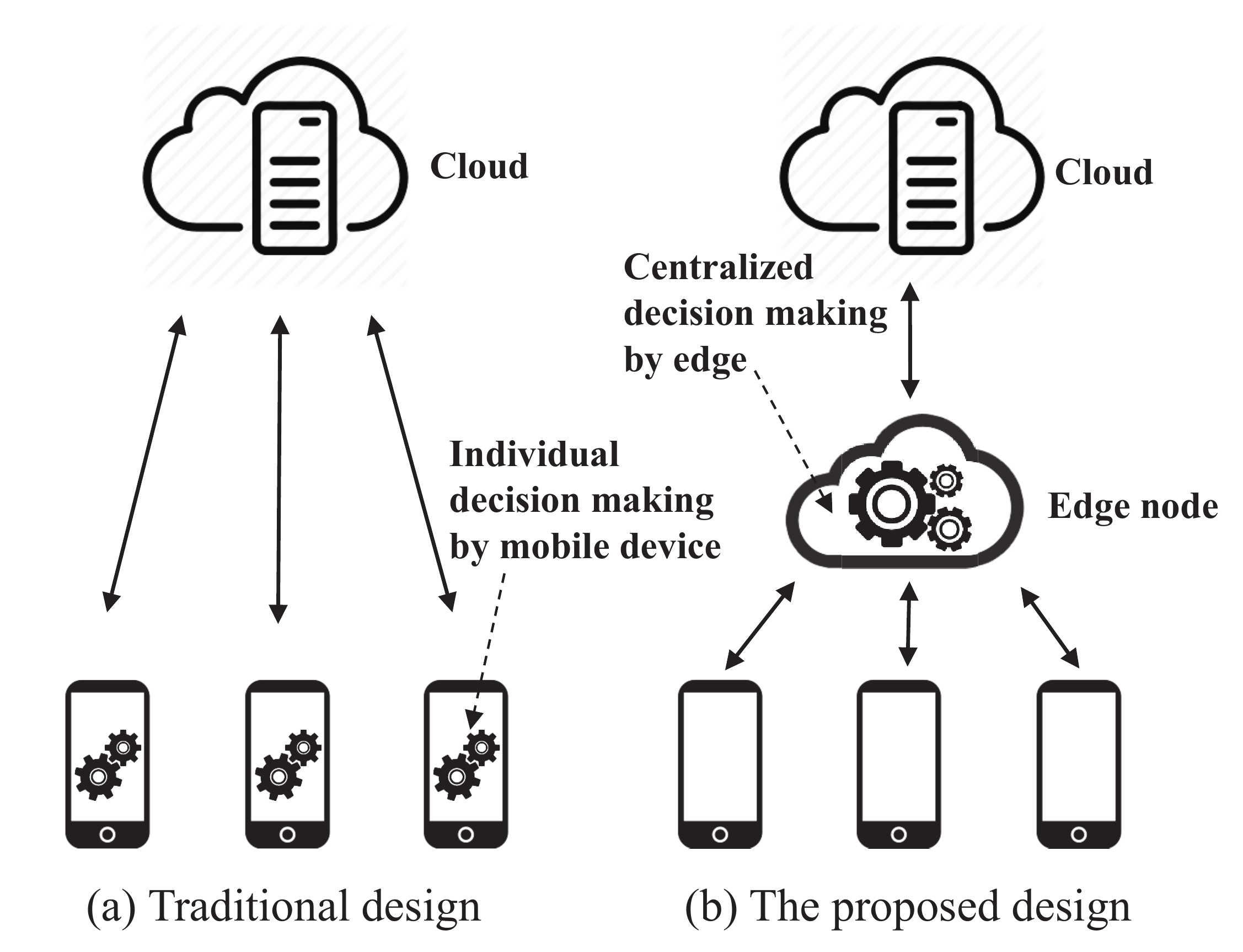}
\caption{Comparison of traditional code offloading and edge-centric code offloading}
\label{fig_arch}
\end{figure}

\begin{table*}[tb]
\newcommand{\tabincell}[2]{\begin{tabular}{@{}#1@{}}#2\end{tabular}}
\centering
\caption{Comparison of offloading systems}
\label{tab:system:comparison}
\begin{tabular} {|p{1.8cm}|p{1.28cm}|p{2.5cm}|p{1cm}|p{1.8cm}|p{1.5cm}|}
\hline
\textbf{\tabincell{l}{Offloading\\systems}} & \textbf{\tabincell{l}{Edge\\supported}} & \textbf{\tabincell{l}{Who makes\\offloading decisions}}  & \textbf{\tabincell{l}{Easy to\\use}} & \textbf{Granularity} & \textbf{\tabincell{l}{QoS\\guarantee}}  \\
\hline
MAUI~\cite{MAUI} & No & Mobile & No & Method-level & No\\
\hline
CloneCloud~\cite{CloneCloud} & No & Mobile & No & Thread-level & No\\
\hline
COMET~\cite{COMET} & No & Mobile & No & Thread-level & No\\
\hline
ThinkAir~\cite{ThinkAir} & No & Mobile  & Yes & Method-level & No\\
\hline
mCloud~\cite{mCloud} & Yes & Mobile  & No & Method-level & No\\
\hline
Echo & Yes & Edge & Yes & Method-level & Yes\\
\hline
\end{tabular}
\end{table*}

The comparison of typical code offloading systems is shown in Table~\ref{tab:system:comparison}. Most of existing code offloading systems~\cite{MAUI, CloneCloud, COMET, ThinkAir} are designed for a two-layer computing hierarchy (i.e., mobile devices and cloud), unaware of the existence of edge nodes. Extending these systems to support edge computing is not an easy task, which involves the redesigning of the whole framework as well as the algorithm for offloading decisions. The first major challenge of building a three-layer, mobile-edge-cloud, code offloading system is to combine the three heterogenous computing platforms and decide on which platform to perform the computation offloading for the best performance improvement. However, the traditional way lets mobile devices individually decide whether a portion of code should be offloaded according to application specifications, the quality of network connection~\cite{DecisionSurvey}, etc., as shown in Fig.~\ref{fig_arch}(a). It is a reasonable design for the traditional two-layer computing hierarchy because cloud has sufficient resources to quickly serve all computing tasks. Since there is little or no contention among tasks at cloud, it is easy for each mobile device to estimate the expected task completion time at cloud, and choose to offload code if it is faster than local execution. Unfortunately, edge nodes have limited resources, and tasks need to contend for running. Without the knowledge of tasks from other devices, it is difficult for a mobile device to estimate the task completion time at edge and make right offloading decisions.

%Most of existing code offloading systems cannot support edge computing because they are designed for a two-layer computing hierarchy (i.e., mobile devices and cloud), unaware of the existence of edge computing nodes. Even though some recent works \cite{RTFace,Glimpse} have made preliminary studies on edge-assist offloading, they target on specific applications with customized runtime, hardly to be extended for other applications. Moreover, they let mobile devices individually decide whether a portion of code should be offloaded to cloud, as shown in Fig. \ref{fig_arch}(a), according the quality of network connection \cite{ThinkAir}. It is a reasonable design for the traditional two-layer computing hierarchy because cloud has sufficient resources to quickly serve all computing tasks. Since there is little or no contention among tasks at cloud, it is easy for each mobile device to estimate the expected task completion time at cloud, and choose to offload code if it is faster than local execution. Unfortunately, edge computing nodes have limited resources, and tasks need to contend for running. Without the knowledge of tasks from other devices, it is difficult for a mobile device to estimate task completion time at edge and make right offloading decisions.

In this paper, we propose an edge-centric code offloading system, called Echo, over a three-layer computing hierarchy consisting of mobile devices, edge and cloud. Echo implements a method-level code offloading on Android-based devices. It allows programmers to use an annotation, i.e., \emph{@Offloadable}, to annotate the methods that are considered to be offloaded. When an annotated method is invoked, the mobile device sends an offloading request to the edge, which then makes centralized offloading decisions, as shown in Fig.~\ref{fig_arch}(b). To optimize the resource usage at the edge, we propose a novel task scheduling algorithm, called Preemption-Constrained Shortest-Remaining-Time-First (PC-SRTF), which aims to minimize the average task completion time without any prior knowledge of future task arrivals. Its basic idea is to let tasks with less remaining time preempt current running tasks, only if the running tasks can finish no later than their local and cloud execution. The decision engine estimates the expected task completion time at edge according to PC-SRTF, compares it with the completion time on local and cloud, and assigns the task to the fastest platform. If a task is decided to be offloaded to the edge, Echo can provide Quality of Service (QoS) guarantee, i.e., even though a task is preempted by future tasks, it can still complete no later than running at mobile device and cloud. For offloaded methods, we further optimize their data uploading process with two mechanisms, lazy object transmission and differential object update, to reduce the amount of data transmitted over the network, so as to reduce the end-to-end delay.

%In addition to the improved performance, the edge interacts end devices and cloud in an asynchronous way, leading to reduced overhead. For example, in traditional code offloading systems, each end device need to measure data transmission delay to cloud before making offloading decisions. In ECHO, edge site can do the measurement on behalf of end devices, and this process does not block the interaction between edge site and end devices.

%Code offloading involves transmitting data from end devices to edge or cloud over the network, which may become a performance bottleneck of code offloading systems. To accelerate data transmission, we propose two mechanisms, lazy object transmission and differential object update, to reduce the amount of data transmitted over the network.

The proposed edge-centric design has three main advantages. First, centralized decision making at edge can provide improved and predictable performance. Second, the edge can interact with mobile devices and cloud in an asynchronous way, leading to reduced overhead. For example, when a method is offloaded to the edge, the edge synchronizes the runtime environment with the mobile device. After that, the edge uploads the same runtime environment to the cloud, so that future offloaded methods at cloud can use it. This process does not block the interaction between the edge and mobile devices. More details can be found in Section~\ref{sec:design}. Finally, Echo can easily deploy new offloading policies by updating the centralized decision engine at edge, without changing the applications on mobile devices.
The main contributions of this paper are summarized as follows.
%\begin{itemize}
%  \item We design and implement a distributed computing framework across mobile device, edge and cloud for code offloading.
%  \item At the edge, we design a centralized decision making algorithm based on PC-SRTF. It can optimize the resource usage at the edge, while guaranteeing offloading performance for each mobile device.
%  \item We enhance Echo's performance by some novel designs, e.g., local code verification, data transmission optimization, and quick service provision at edge.
%  \item A small-scale real deployment  and trace-driven simulations are conducted to evaluate the performance of Echo. The results show that Echo outperforms existing two-layer (mobile device and cloud) code offloading systems by at most 50\% in average task completion time, and more in energy saving.
%\end{itemize}

1) We design and implement a distributed computing framework over mobile devices, edge and cloud for code offloading.

2) At the edge, we design a centralized decision making algorithm based on PC-SRTF. It can optimize the resource usage at the edge, with guaranteed offloading performance for mobile devices.

3) We enhance Echo's performance by some novel designs, e.g., data transmission optimization, and quick service provision at edge.

4) A small-scale real deployment  and trace-driven simulations are conducted to evaluate the performance of Echo. The results show that Echo outperforms existing code offloading systems in both average task completion time and energy consumption of mobile devices.

The rest of this paper is organized as follows. Some important related works are reviewed in Section~\ref{sec:related:work}. System overview is presented in Section~\ref{sec:overview}. Section~\ref{sec:sched} introduces the centralized decision engine at edge, followed by other design details in Section~\ref{sec:design}. Section~\ref{sec:implementation} presents system implementation and Section~\ref{sec:evaluation} shows experimental results. Finally, Section~\ref{sec:conclusion} concludes this paper.

\section{Related Work}
\label{sec:related:work}

Code offloading has attracted many research efforts in recent years~\cite{MAUI, CloneCloud, COMET, ThinkAir, MultisiteOffloading}. Chun et al.~\cite{CloneCloud} have proposed CloneCloud, which can automatically migrate code execution associated with the clone of mobile device's OS to the cloud at thread level. Based on CloneCloud, COMET~\cite{COMET} has implemented code offloading built on distributed shared memory to support multi-threaded applications. However, the implementation of code offloading at the thread level needs customizations of mobile runtime systems, which is intricate and difficult for large-scale deployment.
MAUI~\cite{MAUI} has designed a code offloading framework at method level in order to save the energy consumption of mobile devices. ThinkAir~\cite{ThinkAir} has further optimized resource allocation at cloud, so that it can provide better offloading services to multiple mobile users. Although both MAUI and ThinkAir adopt annotations to implement the method-level offloading, which is similar to Echo, they are not designed to support edge computing. Furthermore, they do not study offloading decisions to guarantee quality-of-services for multiple users, which is one of the main contributions of Echo.

Edge computing~\cite{EdgeComputing} has attracted great interests from both industry and academic~\cite{SmartWorld}. As the network proximity, edge nodes can achieve lower response delay with mobile devices than cloud. Tong et al.~\cite{Tong_2016} have designed a hierarchical edge-cloud architecture to handle offloading requests from mobile devices, and further proposed a workload placement algorithm to maximize utilization of cloud resources. Hou et al.~\cite{HOU2016} have proposed an online algorithm to configure edge-clouds based on history knowledge, which aims to improve the performance of mobile edge computing and minimize the cost. Chen et al.~\cite{CHEN2016} have studied the problem of multi-user code offloading in mobile edge computing, and proposed a distributed offloading decision algorithm based on game theory. Jia et al.~\cite{Jia2015} have proposed an algorithm to properly place edge nodes across wireless metropolitan area network. Above works mainly focus on algorithm design and theoretical analysis for general workloads, lack of implementation and optimization of practical code offloading systems.

Satyanarayanan et al.~\cite{Cloudlet} have first introduced the ``cloudlet", a similar concept with edge nodes. A cloudlet, having good network connection with mobile devices, resembles a small data center deployed at airport, coffee shop, hospital, etc.
Ha et al.~\cite{Gabriel} have built Gabriel for wearable cognitive assistance based on cloudlets. Later, they built OpenStack++~\cite{openstack++} to provide system infrastructure supports. Wang et al. \cite{RTFace} have implemented a system for privacy-aware live video analytics on edge nodes. Glimpse~\cite{Glimpse} is a real-time object recognition system with edge nodes assistance. The above systems focus on specific applications and need great efforts to customize the runtime and edge computing infrastructure. In contrast, Echo is a versatile code offloading solution that can support quick deployment of a wide range of applications.
Zhou et al. have introduced mCloud~\cite{mCloud}, which builds a context-aware offloading framework over client, cloudlet and cloud, but it lets mobile devices individually make offloading decisions, leading to unpreditable offloading performance.

\section{Overview of Echo}
\label{sec:overview}

\begin{figure}
\centering
\includegraphics[width=0.35\textwidth]{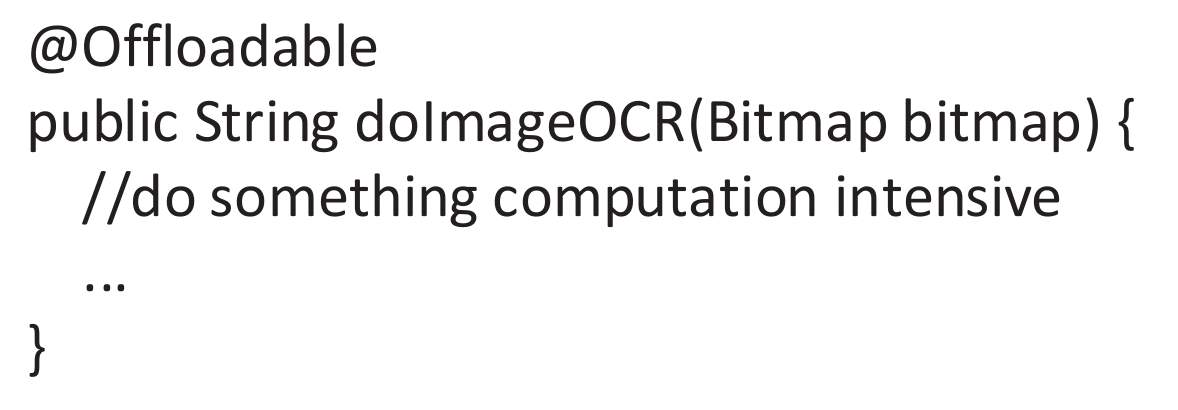}
\caption{An example of method annotation}
\label{fig_annotation}
\end{figure}

\subsection{Problem Statement}
The main goal of Echo is to build a code offloading framework over mobile devices, edge and cloud. It implements a method-level offloading on Android-based devices, without any modifications to mobile operating systems and runtime environment. A critical challenge needs to be addressed by Echo is to decide which method should be offloaded to which computing platform (edge or cloud), so as to accelerate mobile applications.

A typical scenario of using Echo is as follows. Suppose some companies use Echo to develop and deploy mobile applications with offloading capability. Echo provides an annotation-style API, i.e., \emph{@Offloadable}, and programmers use it to annotate methods that are considered to be offloaded, as shown in Fig. \ref{fig_annotation}. This annotation process requires little knowledge on method details, and it works as only a suggestion to Echo's decision engine, which will make final offloading decisions during execution. After annotation, applications are released for downloading and installation. Meanwhile, companies start to provide offloading services by deploying Echo system as well as application code on edge and cloud. They can use public edge and cloud services or deploy their own computing infrastructure. Users whose mobile devices installed with Echo can easily discover available edge nodes and perform code offloading.

%To make ECHO useful and practical, we target on following objectives in its design.
%
%\subsubsection{Superior performance}
%ECHO should fully exploit the benefits of edge sites to outperform existing code offloading systems. Since edge sites are close to mobile devices, they are preferred to be the executors of offloaded methods. However, edge sites have limited hardware resources and can hardly serve too many offloading services. Therefore, it is critical to design a scheduling policy for computing tasks at edge sites.
%%Different from traditional task scheduling within clouds or data centers, ECHO should jointly consider the computing capability of mobile devices, edge sites and cloud, as well as the quality of network connection between them.
%
%\subsubsection{Flexibility in applying different offloading policies}
%Different mobile applications have different offloading policies. For example, some interactive applications desire to quickly obtain the results of offloaded methods, and a policy with the objective of minimizing response time should be applied. For some delay-tolerant applications focusing on energy saving of end devices, ECHO should apply a policy that offloads as many as methods to edge sites or cloud.
%
%\subsubsection{Quick deployment}

\subsection{Architecture Overview}
% describe the workflow

\begin{figure}
\centering
\includegraphics[width=0.48\textwidth]{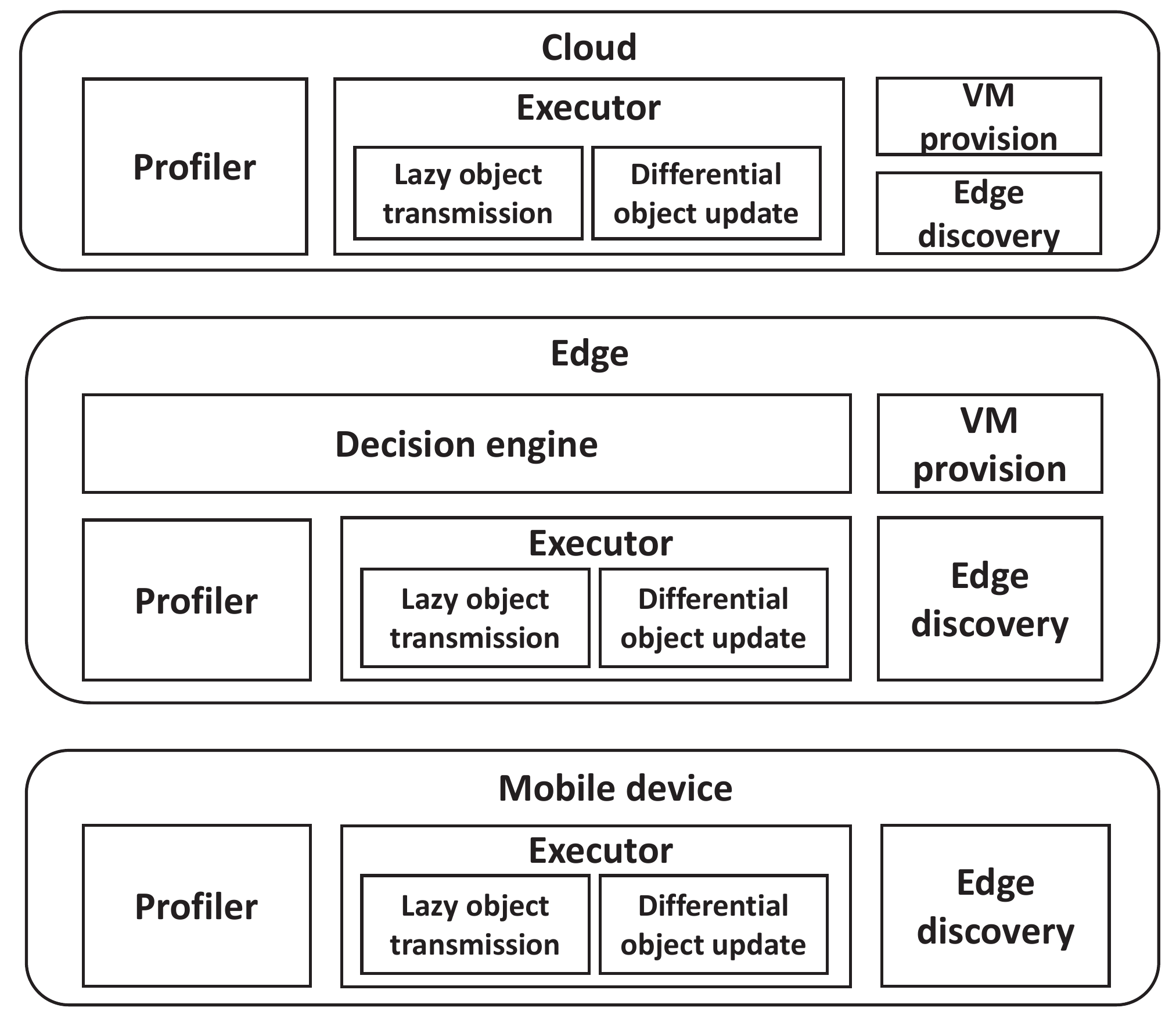}
\caption{Echo architecture.}
\label{fig_framework}
\vspace{-1.5em}
\end{figure}

The Echo architecture is shown in Fig. \ref{fig_framework}. Edge discovery enables a mobile device to find an appropriate edge node for code offloading. Then when an annotated method is invoked in mobile devices, it is first checked by a module called code profiler included in the profiler. The code profiler will analyze if the method contains unoffloadable operations, e.g. user input or reading sensors (e.g., camera, accelerometer and gyroscope). With this analysis, it removes the necessary of examining the code line by line for programmers.

%If this method contains local operations, e.g., user input or reading sensors (e.g., camera, accelerometer and gyroscope), it should not be offloaded. Code verification is necessary because programmers annotate methods using ``@Offloadable'' according to their experiences, without checking the code line by line. This module can quickly filter the methods that cannot be offloaded.

If a method passes the code inspection, the mobile device sends an offloading request to the centralized decision engine at the edge. This kind of method is also referred to a task in the following presentation. The main function of decision engine is to estimate the expected task completion time on mobile device, edge and cloud, respectively, and assign the task to the computing platform (mobile device, edge or cloud) with shortest task completion time. Since there is little or no resource contention at cloud, the task completion time can be calculated by profiling task running time, data uploading and downloading time, and summing them all simply. However, the expected task completion time at edge depends on the task scheduling algorithm. In this paper, we propose a Preemption-Constrained Shortest-Remaining-Time-First (PC-SRTF) algorithm to minimize the average task completion time of tasks offloaded to edge while avoiding starvation, without any prior knowledge of future task arrivals. According to PC-SRTF, the decision engine estimates the expected task completion time at the edge, and makes a final offloading decision that is sent back to the mobile device. Specifically, it can provide QoS guarantee when offloading performing at the edge.

If an method is decided to be offloaded to the edge, Echo quickly prepares an executor based on a virtual machine (VM) with configured runtime at the edge, and synchronizes its status with the one at the mobile device. Meanwhile, with the method execution, we also need to upload input data and objects involved in this method. To optimize this synchronization process, we propose a mechanism called lazy object transmission, which synchronizes an object only when it is used during execution. Furthermore, Echo caches transmitted objects at the mobile and edge and adopts the differential object update. When the same object is synchronized again, Echo compares it with the cached one and transmits only the different parts. The above techniques are also applied on executors on cloud.

\section{Centralized Decision Engine at Edge}
\label{sec:sched}

The decision engine at the edge is the most critical module of Echo to provide offloading QoS guarantee. We design the decision engine with the objective of minimizing average task completion time, without prior information of future offloading requets. We first present the system model, and then describe the decision-making algorithm based on PC-SRTF.

\subsection{System Model}
We consider an edge node with limited hardware resources, which are virtualized as a set $M$ of identical virtual machines (VMs). Offloading tasks generated by mobile devices arrive in an online manner, i.e., the edge has no knowledge about future task arrivals. For each task $i$, its completion time $T(i)$ is defined as follows.

\begin{equation}
\label{eq_time}
T(i) = \left\{
				\begin{array} {l}
						R_{m}(i), \text{mobile device}; \\
						D_{c}^{up}(i)+R_{c}(i)+D_{c}^{down}(i), \text{cloud}; \\
						D_{e}^{up}(i)+W_{e}(i) + R_{e}(i)+D_{e}^{down}(i), \text{edge}.
				\end{array}
\right.
\end{equation}

Case 1: task $i$ runs at the mobile device. The task completion time $T(i)$ is equal to the local running time $R_{m}(i)$, which is determined by the task size and the hardware of the mobile device.

Case 2: task $i$ is offloaded to the cloud. The mobile device needs to upload data to cloud, and the uploading time is denoted by $D_{c}^{up}(i)$. We suppose the cloud has sufficient resources, and the task can be immediately served by $R_{c}(i)$ time. After execution, the mobile device downloads results and the downloading time is denoted by $D_{c}^{down}(i)$.

Case 3: task $i$ is offloaded to the edge. The data uploading and downloading time are denoted by $D_{e}^{up}(i)$ and $D_{e}^{down}(i)$, respectively. Different from the cloud, an edge node has limited resources and the task may need to wait before being served due to resource contention. Therefore, the time spent at edge consists of task waiting time $W_{e}(i)$ and running time $R_{e}(i)$. The task waiting time $W_{e}(i)$ depends on the task scheduling algorithm adopted by edge.

The profiling modules, deployed at mobile devices, edge and cloud, collect the information of task size, VM power of different computing platforms and quality of network connection (details in Section~\ref{subsec:profiling}), so that the decision engine can accurately estimate task running time ($R_{m}(i)$, $R_{e}(i)$ and $R_{c}(i)$), data uploading time ($D_{e}^{up}(i)$ and $D_{c}^{up}(i)$) and downloading time ($D_{e}^{down}(i)$ and $D_{c}^{down}(i)$). Since edge is located closer to mobile devices than cloud, we have $D_{e}^{up}(i)<D_{c}^{up}(i)$ and $D_{e}^{down}(i)<D_{c}^{down}(i)$. Moreover, we usually have $R_{c}(i)\leq R_{e}(i)< R_{m}(i)$ because edge and cloud have powerful hardware. Note that tasks have different characteristics on data transmission and task execution. For example, some tasks with little computation need to upload a large amount of data, whereas others (e.g., compute-intensive tasks) upload small data, but the execution is time-consuming.

\subsection{Decision Making based on PC-SRTF}
For each offloading request $i$, the decision engine estimates the expected task completion time at mobile device, edge and cloud, respectively, according to (\ref{eq_time}). It then makes an offloading decision by assigning the task to the platform with minimum completion time.

We let $T_{m}(i)$ and $T_{c}(i)$ denote the completion time of task $i$ at mobile device and cloud, respectively.
Based on measurements by the profiler, we can easily calculate $T_{m}(i)$ and $T_{c}(i)$.
Next, we focus on task scheduling at edge, so that we can calculate the expected task completion time at edge. Given a number of tasks offloaded to the edge, to minimize average task completion time, an intuitive design is to always schedule the task with minimum remaining time. Unfortunately, this simple heuristic would lead to starvation for long tasks if small tasks frequently arrive. When a task is offloaded to the edge, its completion time should be no later than a deadline $H(j)=\min\{T_{m}(i), T_{c}(i)\}$. Otherwise, this task should be assigned to the cloud or the mobile device. This observation motivates us to design an algorithm called Preemption-Constrained Shortest-Remaining-Time-First (PC-SRTF), which allows smaller tasks to preempt the execution of longer tasks, only if the preemption makes longer tasks complete no later than their deadlines.

The pseudo code of the decision-making algorithm based on PC-SRTF is shown in Algorithm \ref{alg1}. For each VM $m_{q}\in M$, we maintain a queue $Q_{q}$ for waiting tasks. The VM $m_{q}$ sequentially executes the tasks in $Q_{q}$. When a new task $i$ arrives at time $t$, we calculate its expected completion time at edge from line 2 to 24. Since task $i$ is not really scheduled during this stage, we define a temporary queue $Q'_{q}=Q_{q}$ in line 4 and do the following calculation based on it.

\begin{figure}
\centering
\includegraphics[width=0.4\textwidth]{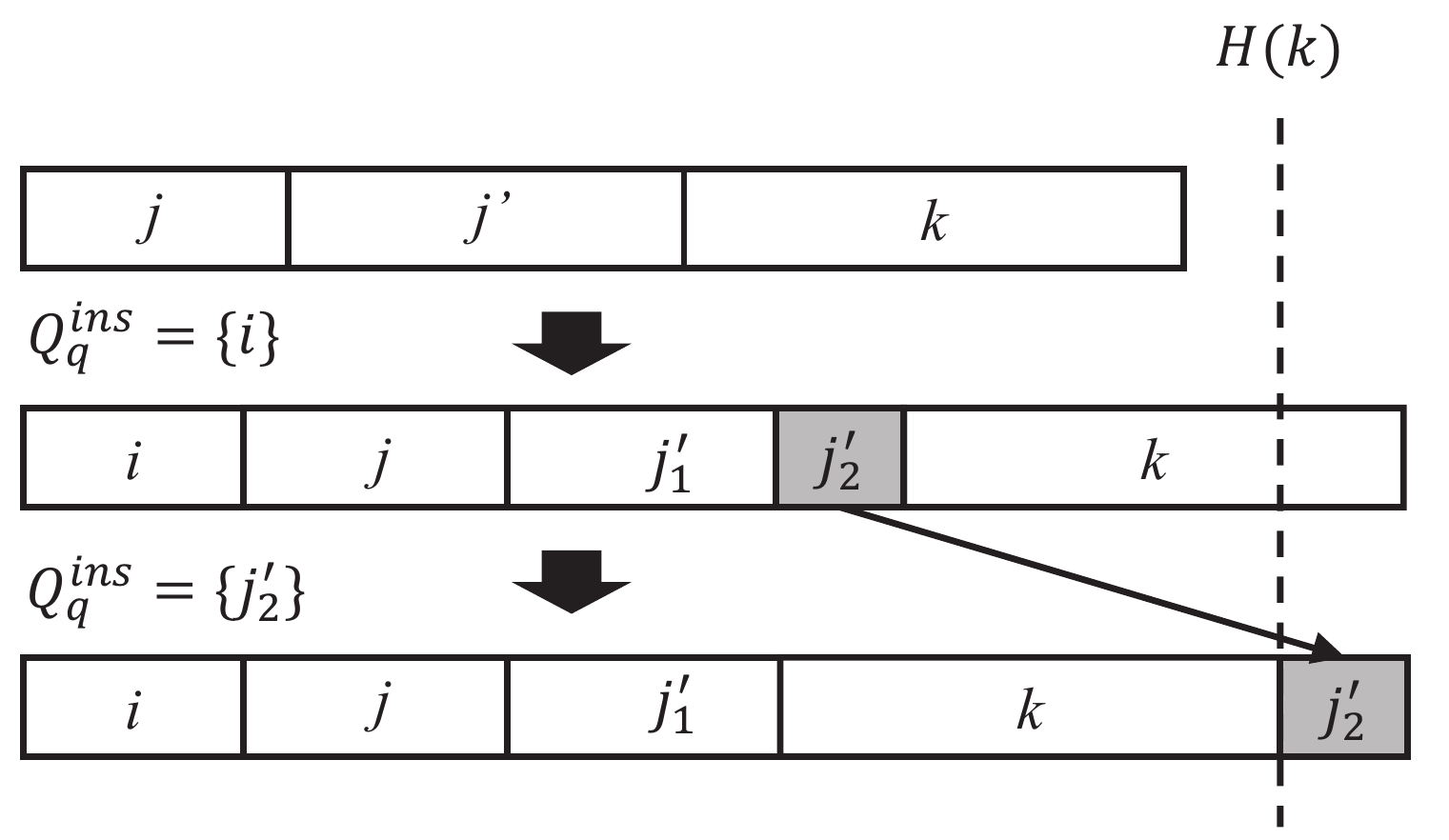}
\caption{An example of task preemption}
\label{fig_violation}
\end{figure}

We search queue $Q'_{q}$ to find the first task $j$ whose remaining time is greater than task $i$. If such a kind of task $j$ cannot be found, we put task $i$ at the tail of $Q'_{q}$. Otherwise, a queue $Q_{q}^{ins}$ is defined to include the tasks that will be inserted into $Q'_{q}$, and it is initialized as $\langle i\rangle$. We let $p$ point to the header of task $j$, where we will insert $Q_{q}^{ins}$.
In the following \textbf{while} loop from line 11 to 20, we insert $Q_{q}^{ins}$ at the place pointed by $p$. If any task $k$ completes later than its deadline $H(k)$ due to this insertion, we need to adjust the scheduling by postponing $k$ until it can finish at $H(k)$. There must be a portion of workloads evicted from $Q'_{q}$, and we use them to replace the contents in $Q_{q}^{ins}$, so that they will be inserted after task $k$ in the next iteration. This process can be illustrated by the example shown in Fig. \ref{fig_violation}. Suppose there are three unfinished tasks in $Q'_{q}$ and a new arrived task $i$ is inserted into the header of $Q'_{q}$ because its smallest remaining time. We postpone the task $k$ until $H(k)$, and a portion of task $j'$, which is denoted by $j'_{2}$, is evicted from the scheduling. The evicted workload $j'_{2}$ is moved into $Q_{q}^{ins}$, which will be considered for insertion in the next iteration.

\begin{algorithm}[t]
\caption{\label{alg1}Decision Making Algorithm based on PC-SRTF}
\begin{algorithmic}[1]
\STATE Maintain a task queue $Q_{q}$ for each VM $m_{q}\in M$;
\FOR {each new task $i$ arriving at time $t$}
	\FOR{each VM $m_{q}$}
		\STATE $Q'_{q}=Q_{q}$;
		\STATE Search queue $Q'_{q}$ from the beginning, and find the first task $j$ whose remaining time is greater than task $i$;
		\IF{cannot find such a task $j$}
			\STATE Put task $i$ at the tail of $Q'_{q}$;
		\ELSE
			\STATE $Q_{q}^{ins}=\langle i\rangle$;
			\STATE $p \to$ the header of task $j$;
			\WHILE {$p$ is not NULL}
				\STATE Insert $Q_{q}^{ins}$ at $p$;
				\IF {any task $k$ completes later than $H(k)$}
					\STATE Postpone task $k$ until it can finish at $H(k)$;
					\STATE A portion of workloads before $k$ are evicted, and use them to replace the contents in $Q_{q}^{ins}$;
					\STATE $p\to$ the tail of task $k$;
				\ELSE
					\STATE $p\to NULL$;
				\ENDIF
			\ENDWHILE
		\ENDIF
		\STATE Calculate the completion time growth $\Delta T_{q}$ of tasks in $Q'_{q}$;
%		\STATE According to the updated $Q_{q}^{j}$, calculate response time growth $g_{q}$ of all tasks in $Q_{q}^{j}$, and task $i$'s expected response time $T_{e}^{q}(i)$.
	\ENDFOR
	\STATE $q^{*}=\arg\min\{\Delta T_{q}\}$;
    \STATE Compare $T_{e}^{q^{*}}(i)$, $T_{c}(i)$, and $T_{m}(i)$, and assign the task $i$ to the platform with minimum completion time;
	\IF{task $i$ should be offloaded to the edge}
		\STATE Assign task $i$ to $m_{q^{*}}$ and replace queue $Q_{q^{*}}$ with $Q'_{q^{*}}$;
		\STATE $H(i)=t+\min\{T_{c}(i), T_{e}(i)\}$;
	\ENDIF
\ENDFOR
\end{algorithmic}
\end{algorithm}

After that, we find the VM $m_{q^{*}}$ with the minimum expected completion time growth. The expected completion time of task $i$ on this VM is denoted by $T_{e}^{q^{*}}(i)$. We compare $T_{e}^{q^{*}}(i)$ with $T_{c}(i)$ and $T_{m}(i)$, and assign task $i$ to the platform with minimum completion time. If task $i$ should be offloaded to the edge, i.e., $T_{e}^{q^{*}}(i)<T_{c}(i)$ and $T_{e}^{q^{*}}(i)<T_{m}(i)$, we assign task $i$ to VM $m_{q^{*}}$ and update its scheduling according to $Q'_{q}$.

%Since tasks before $j$ are not affected by this insertion, we focus on task $j$ and its successors, which are maintained in a subqueue $Q_{q}^{j}$, as shown in line 5. Next, we sequentially check tasks in $Q_{q}^{j}$ in the for loop from line 6 to 11. If a violation happens for task $k$, we postpone task $k$ until it can finish at $H(k)$. Some portion cannot be accommodated before task $k$ is moved after $k$. We use the example in Fig. \ref{fig_violation} to illustrate this process. Suppose there are three unfinished tasks in $Q_{q}^{j}$ and the new arrived task $i$ is inserted into the header of $Q_{q}^{j}$ because its smallest remaining time. We postpone the task $k$ until $H(k)$, and schedule part of task $j'$ after $k$. Similar operations are applied to successive tasks in $Q_{q}^{j}$. Finally, we choose VM $m_{q'}$ leading to the minimum response time growth and compare it to $T_{c}(i)$ and $T_{e}(i)$ to make final offloading decisions.

\begin{figure}[t]
\centering
\includegraphics[width=0.45\textwidth]{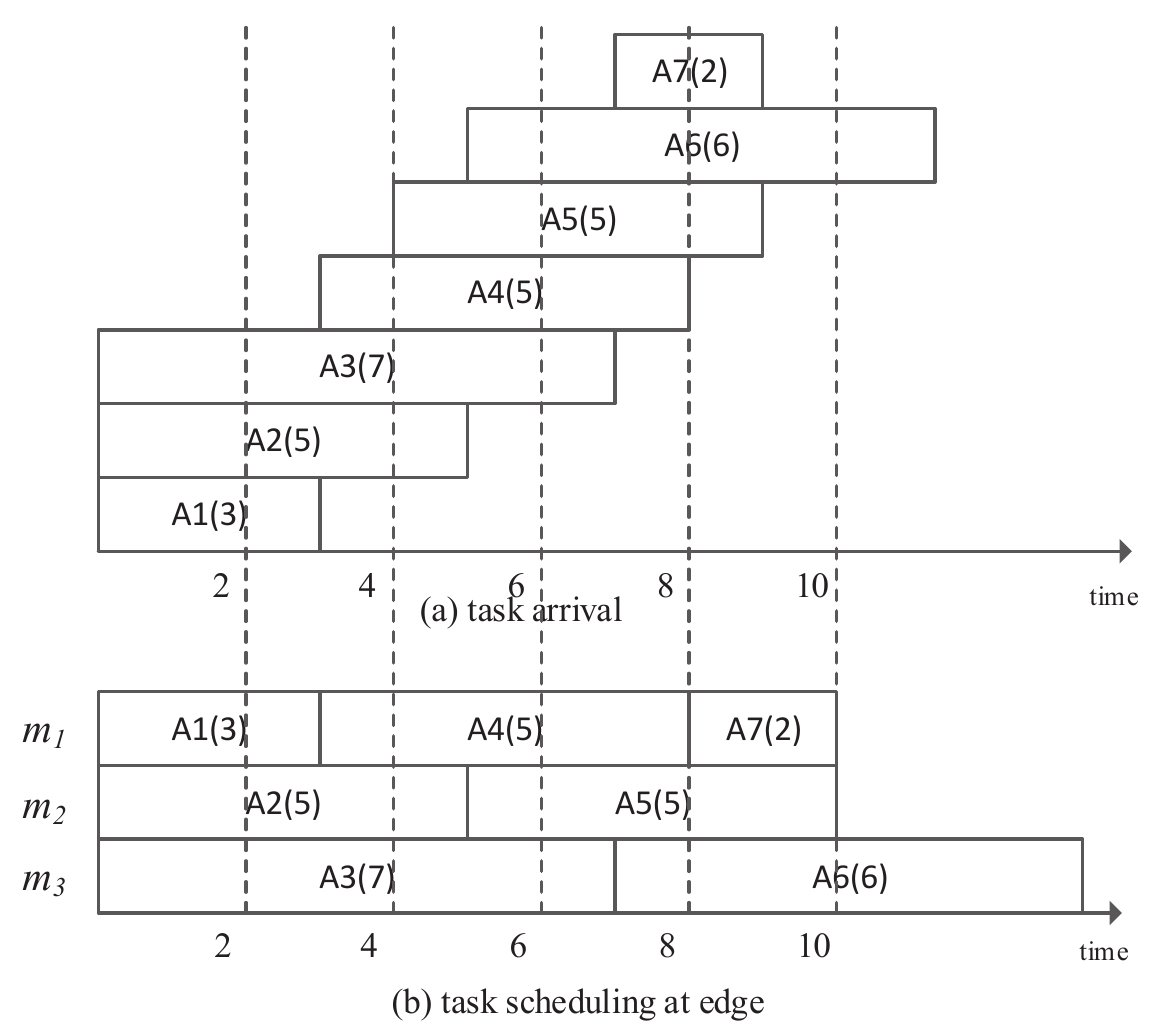}
\caption{An example of PC-SRTF}
\label{fig_example}
\end{figure}

For easy understanding, we use an example in Fig. \ref{fig_example} to show how the proposed algorithm works. There are three virtual machines available at the edge. In the beginning, three tasks $A_{1}$, $A_{2}$ and $A_{3}$ are ready for running, and they are assigned to three virtual machines, respectively. The number in the bracket indicates the task size.
%For all tasks in this example, we further assume that their completion time on mobile device or cloud is longer than the time on edge even considering the task waiting time.
Then, task $A_{4}$ with  size of 5 arrives at the 3rd time slot, but it cannot preempt any running tasks because their remaining time is less than $A_{4}$. We choose to schedule $A_{4}$ after $A_{1}$ on $m_{1}$ because it leads to the minimum growth of average completion time. Later, tasks $A_{5}$ and $A_{6}$ arrive, and it is easy to see that they should be scheduled on $m_{2}$ and $m_{3}$, respectively. At the 7th time slot, task $A_{7}$ arrives. It cannot preempt $A_{4}$ on $m_{1}$ because $A_{4}$'s remaining time is not more than $A_{7}$. The completion time growth is 10 if $A_{7}$ is scheduled after $A_{4}$. On $m_{2}$, we suppose that $A_{5}$'s deadline is the 11th time slot, and $A_{7}$ can preempt $A_{5}$ only at the 7th time slot, leading to the growth of 13 in average completion time. On $m_{3}$, $A_{7}$ can run after $A_{3}$ and the growth is 11. Therefore, we finally decide to schedule $A_{7}$ after $A_{4}$ on $m_{1}$.

\section{Design Details}
\label{sec:design}

Echo is designed as an integrated offloading system, and we develop functional modules to guarantee its QoS and improve the overall system performance. The details of these modules are presented in this section.

\subsection{Edge Discovery}
%The first step of edge-assist offloading is the edge discovery. In indoor scenarios like laboratory, office and coffee shop, edge nodes are usually deployed nearby mobile devices and they share the same network domain. In this case, it is natural to use local area network discovery protocols (e.g., zero-configuration) to find edge nodes. However, if people outdoor search for the offloading services, it is infeasible to use the way of local resource discovery. In Echo, we develop a two-level edge discovery mechanism: a local discovery and a global one.

The first step of edge-assist offloading is the edge discovery. According to the different types of network connections between mobiles and edges, we develop a two-level edge discovery mechanism: a local discovery and a global one. The edge discovery modules, deployed on mobile, edge and cloud, work collaboratively to perform the two-level discovery.

The local edge discovery uses the zero-configuration protocol (Bonjour protocol\footnote{https://en.wikipedia.org/wiki/Bonjour\_(software)}) to find available edge nodes that are in the same network domain (i.e., LAN) with mobile devices. In the process of local discovery, an edge node announces its offloading service by network broadcast, which includes the message of service name, IP address and port number. An example of this broadcast is \_echo.$<$10.136.3.71$>$.$<$8022$>$. A mobile device searching for this type of service (e.g., echo service) receives the broadcast and negotiates with the edge, then finally connects to the edge with the specific port.
%The local discovery achieves a fast edge computing service discovery but limited in local area network as routers filter these multicast messages.

The global discovery uses cloud to help discovery. The cloud acts as a directory server, which holds items of all available edge nodes geographically dispersed. Edges register to cloud with the information of IP address, location (longitude and latitude) and resource availability. Then they continuously send heartbeat messages indicating their status (active or disconnected). A mobile device asking for offloading service sends cloud a request, which includes user id, application id, IP address and location of the mobile device. To respond this query, cloud searches its database and finds an edge that is close to the mobile device according to their location proximity. Finally, the mobile device connects to the edge specified by the cloud.
\subsection{Echo Profiler}
\label{subsec:profiling}
Echo's edge-centric decision is based on the estimation of application execution time at mobile, edge, and cloud, respectively. For a precise estimation, we use the way of history-based profiling predication, which has been demonstrated by previous works~\cite{HistoryPrediction, MAUI}. To build the predication model, we collect resource information and code behavior by profilers, which include hardware profiler, network profiler and code profiler, deployed on mobile, edge and cloud, shown in Fig.~\ref{fig_framework}. Hardware profiler provides hardware state about the power of CPU, memory etc., with respect to virtual machines at edge and cloud or mobile devices. Network profiler probes network characteristics, such as network type, bandwidth and network latency. Code profiler observes the program behavior when the program runs with different input data, and it records the size of input data, the number of method instruments and the overall of method execution time. Each time when a method is offloaded, all the above information is combined as a running log. Then, these history logs feed the predication model, which can estimate the execution time of an offloading method at runtime.
\subsubsection{Code Profiler}
\label{subsec_code_verification}
\begin{figure}[tb]
\centering
\includegraphics[width=0.40\textwidth]{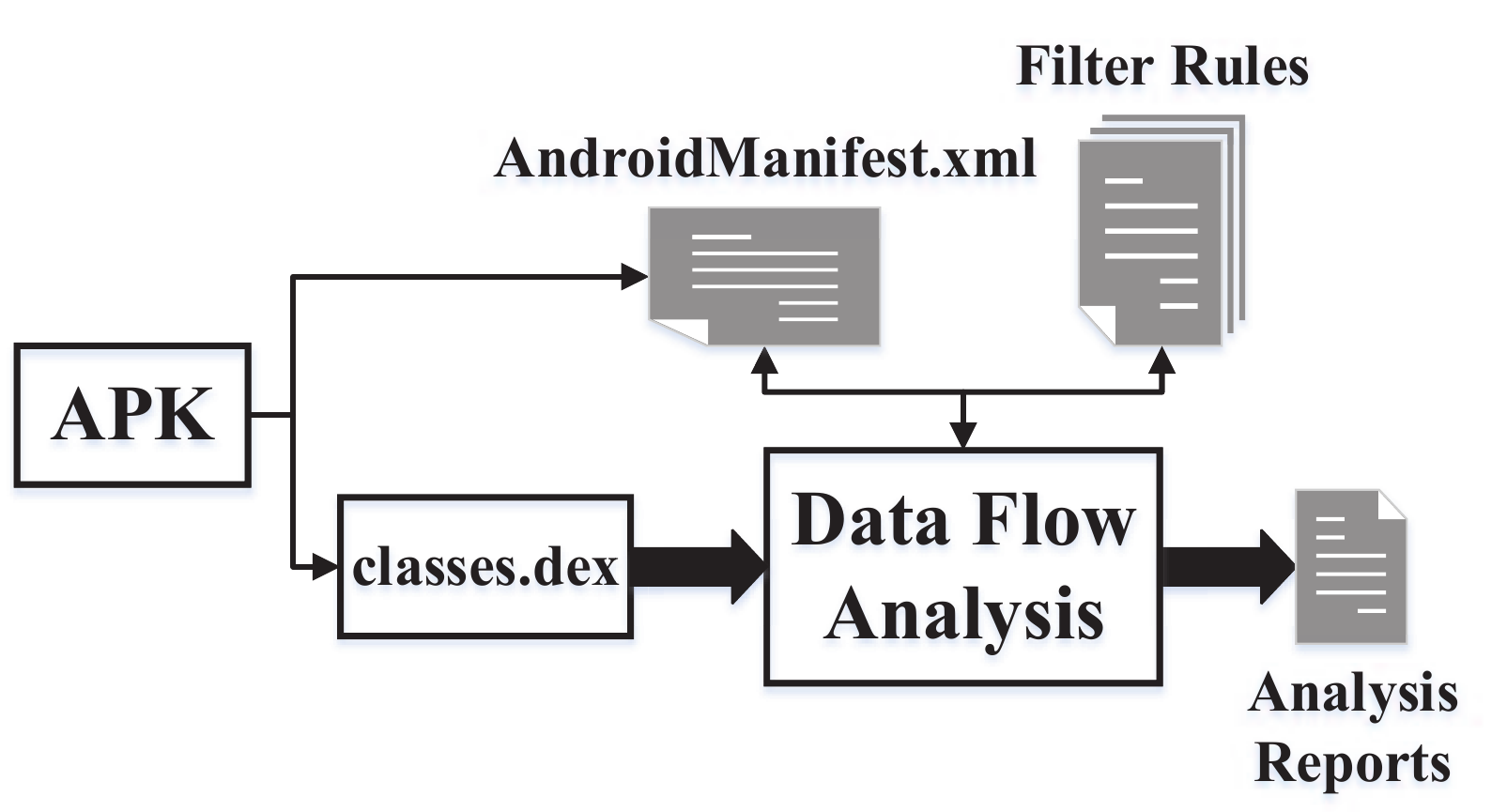}
\caption{Flowchart of code analysis} \label{fig_code_verification}
\end{figure}

%Since not all annotated methods can be offloaded, we develop a code verification module to filter the methods involving hardware control, UI manipulation or other unoffloadable operations. The code verification module can be used as a static code inspection tool that helps programmers quickly checking which code can be offloaded.
As discussed above, the code profiler measures the running data of method execution. Besides, it provides a feature of code inspection on annotated methods, to check if those methods can be offloaded. We build call graphs~\cite{FlowDroid} to analyze if the annotated methods contain unoffloadable operations, e.g., hardware control, UI manipulation, etc., and those methods containing these operations should be anchored at mobile. The process of code analysis is shown in Fig.~\ref{fig_code_verification}, and the analysis module is designed for Android program. In the beginning, it takes an APK file (Android application zip file) as input, and analyzes the AndroidManifest.xml file to get the main entrance of the whole application. It then performs a rule-based dataflow analysis. In our current implementation, the following rules are used.

\textbf{\emph{Rule 1}}. \emph{Mobile hardware operations must be anchored on mobile devices}. These operations involve using hardware sensors like camera, GPS, microphone, etc.

\textbf{\emph{Rule 2}}. \emph{UI manipulations must be anchored on mobile devices}. It means all operations related to Android view manipulations need to be executed on mobile devices.

\textbf{\emph{Rule 3}}. \emph{IO (Input/Output) operations must be anchored on mobile devices}. This rule filters the read and write operations of IO streaming including file IO and network IO. It also limits data storage operations like using SQLite database or SharedPreference\footnote{https://developer.android.com/training/data-storage/shared-preferences.html}.

\textbf{\emph{Rule 4}}. \emph{Graphics rendering and display must be anchored on mobile devices}. Android offers a variety of graphics rendering APIs, and these APIs interact with mobile graphic drivers which are hardware-dependent.

\subsection{Data Transmission Optimization}

When Echo decides to offload a method, it synchronizes the executor status by uploading related objects from mobile devices to the edge or cloud. We design two mechanisms to accelerate this synchronization process.

\subsubsection{Lazy object transmission}
\label{sec:lazy:transmission}
In an intuitive design, all objects associated with an offloaded method should be transmitted to the edge or cloud. However, by carefully examining the code execution process, we find that not all these objects will be used. This observation motivates us to design a lazy object transmission mechanism~\cite{CopyByReference}. Specifically, when a method is offloaded, we create proxies for associated objects at edge or cloud, instead of transmitting real ones. Once an object is referred, an object transmitting request is sent back to the mobile device, which then transmits the real one. As a result, we remove unnecessary object transmission.

\begin{figure}[tb]
\centering
\includegraphics[width=0.4\textwidth]{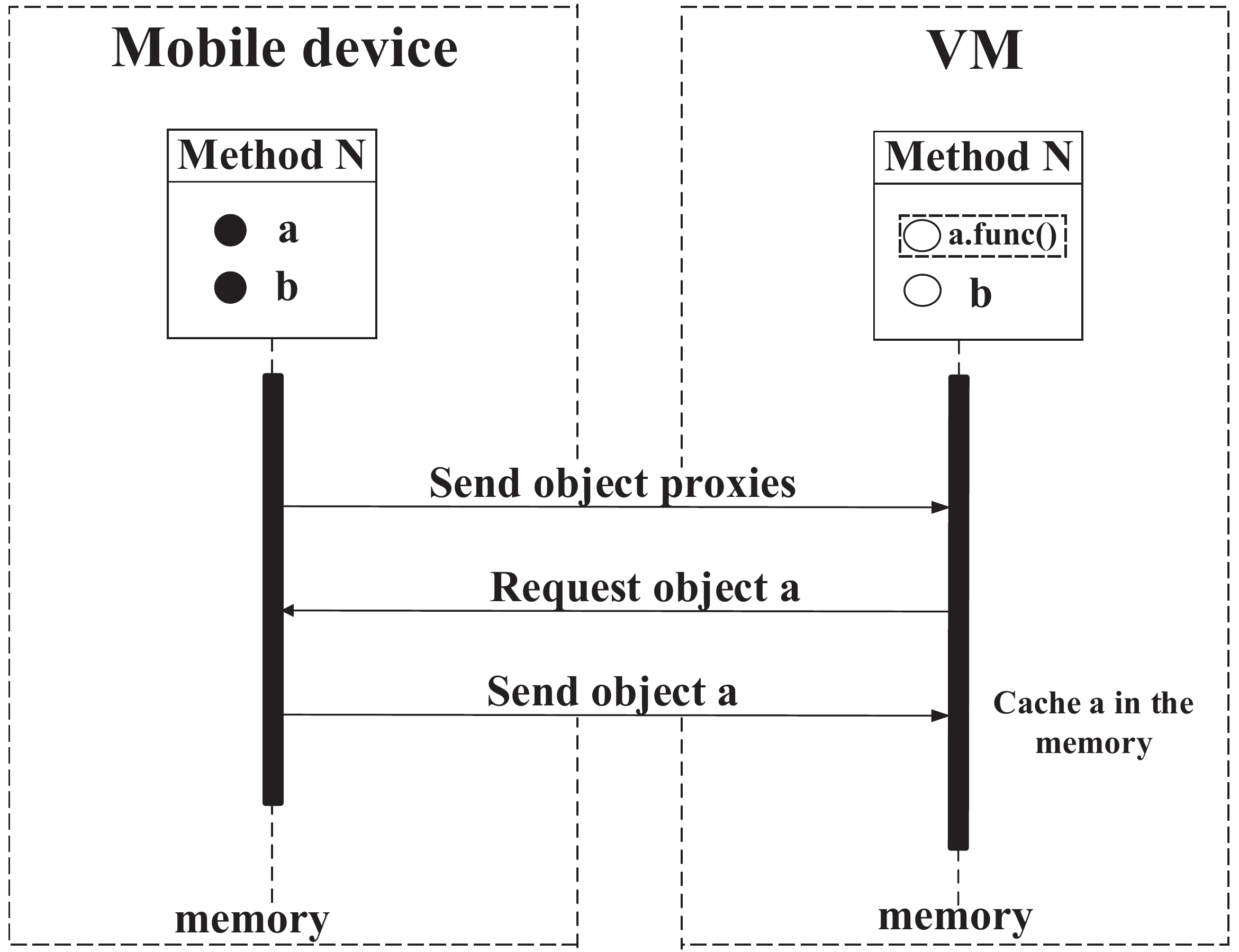}
\caption{An example of lazy object transmission}
\label{fig_proxy_detail}
\end{figure}

Fig.~\ref{fig_proxy_detail} illustrates an example of lazy object transmission, where an offloading method \emph{N} involves two objects: \emph{a} and \emph{b}. In the beginning, the mobile device only sends object proxies of small size to the VM on the edge or cloud. Once the object \emph{a} is referred, e.g., its method is invoked, the mobile device then transmits the object \emph{a}. In contrast, the object \emph{b} is not transmitted because it is not referred.
%COSA instantiates a proxy object \emph{p}, which contains a light object \emph{n} including empty \emph{a} and \emph{b}; meanwhile, it stores the real object on its local cache. When the member \emph{a} is actually used at the server, as shown in the picture, it invokes method \emph{func}. Then, COSA will perform a real object loading; first it searches the cache of the server side, and loads the object if cache hit otherwise asks for it from the mobile side. The mechanism initializes the minimum data of object and provides a lazy object transmission to remove unnecessary data transmission, like member \emph{b} in this example. We find this object proxy is time and bandwidth saving in practice.

\subsubsection{Differential object update}
\label{sec:differential:update}
We also observe that when a mobile device continuously offloads methods to the edge or cloud, and these methods usually refer to some common objects (e.g., shared buffers or resource objects). In order to reduce the amount of data transmission, we propose to only update the differential parts of these objects during synchronization~\cite{DeltaCompress}. Specifically, when an object is used by an offloaded method for the first time, it is cached at both local and remote. When this object is used again, we compare it with the cached one on the mobile device, then only the different parts are transmitted. Especially, when the object is reused without changed, none object data needs to be transmitted. Furthermore, when an object at edge (cloud) is updated, we also synchronize it to cloud (edge), so that later offloaded methods at cloud (edge) can reuse this object. The object synchronization between edge and cloud is conducted in an asynchronous way, without affecting the interaction between mobile devices and edge.

\subsection{Quick Service Provision}
\label{sec:service:provision}
%VM-based code offloading can provide a consistent environment for code execution across heterogeneous computing platforms. It has been widely adopted by existing works, e.g., \cite{CloneCloud,ThinkAir}, which start a dedicated VM for each mobile device requesting offloading service. Unfortunately, the edge can support only a limited number of simultaneously running VMs. When a busy edge computing node receives requests from a new mobile device, we may need to release resources by stopping a running VM, so that we can start a new one. This process incurs large overhead, especially when many mobile devices contend for edge resources. To address this challenge, we propose to reuse existing VMs by setting up the execution environment for different users. Specifically, we pre-install the applications with offloading capability on these VMs. When an offloaded method is assigned to a VM, Echo loads related objects and schedules the method for running. In addition, Echo periodically removes some objects that have not been referred for a long time.

VM-based code offloading can provide a consistent runtime for code execution across heterogeneous computing platforms. It has been widely adopted by existing works, e.g.,~\cite{MAUI} and~\cite{CloneCloud}, which start a dedicated VM for each mobile device requesting offloading service. The dedicated VM is usually created by launching a base VM that then synthesizes a differential VM binary, which is the different part of VM between the base one and the mobile device's specific VM, and this process is called VM synthesis~\cite{Justintime,OnDemandProvisioning}. This way is easy to be deployed, but resource-intensive; moreover, mobile devices should transmit their VM binary files to the server, which is time-consuming. Conversely, in Echo, for quick offloading service provision, we propose a lightweight VM synthesis by reusing existing VMs and setting up the execution environment for different users. To provide an executable VM for a mobile device, we launch a base VM and install application packages associated with configurations for the specific mobile device. Furthermore, when a mobile device searches offloading service for a specific application at edge, we first choose the VMs that have installed the application package with corresponding program's data. Therefore, the edge node can quickly respond to the mobile device without heavy data transmission.

\section{Implementation}
\label{sec:implementation}

%\begin{figure}
%\centering
%\includegraphics[width=0.4\textwidth]{figs/executionhandler.eps}
%\caption{The execution of code offloading between mobile devices and cloudlets/cloud by execution handler and offloading executor}
%\label{fig_exeution_handler}
%\end{figure}

\begin{figure}[tb]
\centering
\includegraphics[width=0.35\textwidth]{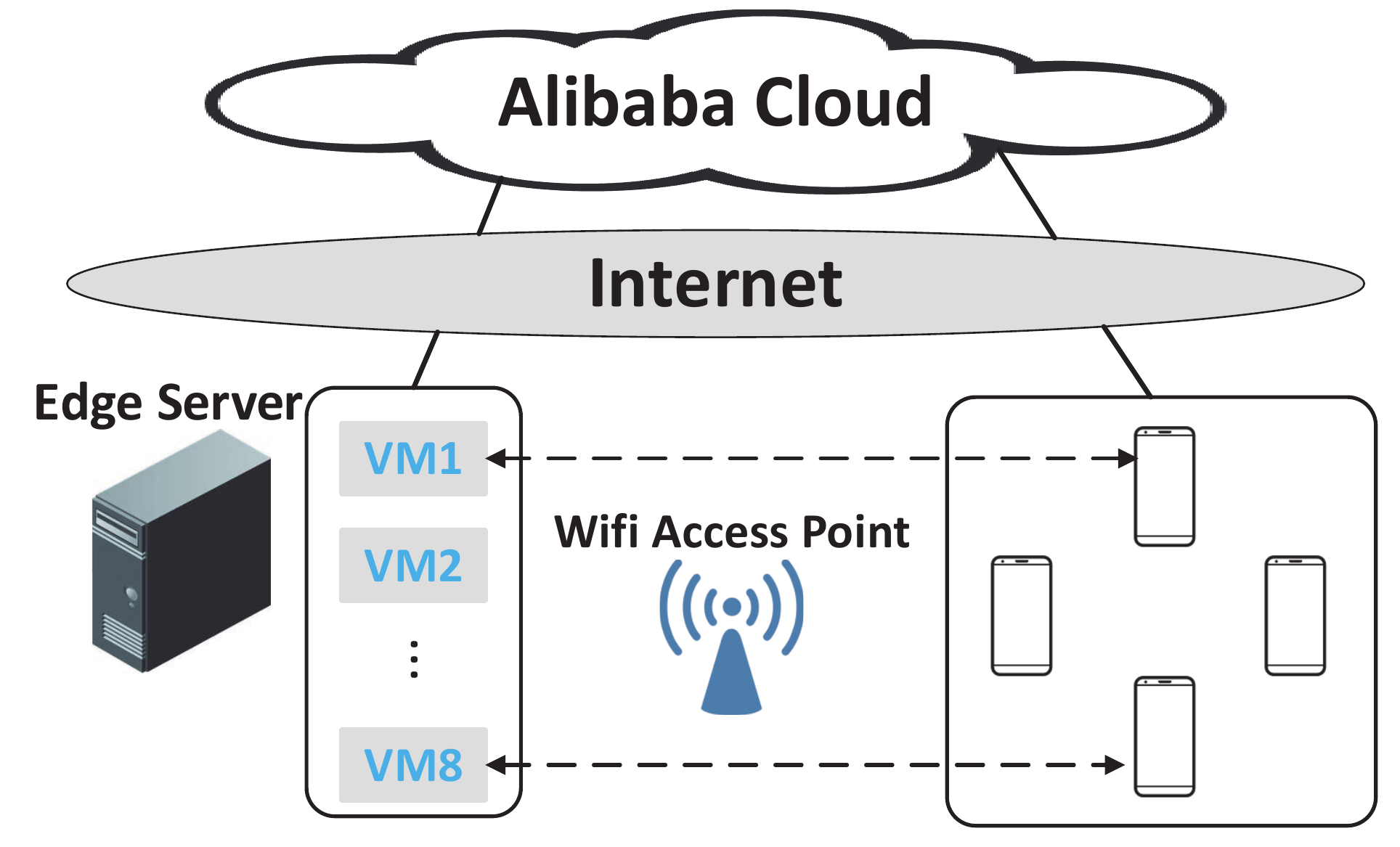}
\caption{Deployment of testbed}
\label{fig:testbed}
\end{figure}

We implement Echo using Java for Android mobile devices. It uses the paradigm of \emph{aspect-oriented programming (AOP)}\footnote{https://en.wikipedia.org/wiki/Aspect-oriented\_programming}, which allows us to insert specific offloading operations into source code. To integrate Echo with Android applications, we use \emph{aspectj}\footnote{https://eclipse.org/aspectj/}, which can recompile Java source code to enable code offloading. We use \emph{JSON}\footnote{http://www.json.org/} for object serialization because it is more versatile and effective compared to Java native serialization. Offloaded methods are executed at edge or cloud by using \emph{Java Reflection}. The execution environment at the edge is built on the virtual machine of Android-x86, which can run Android applications on x86 platforms.

The code analysis module of Echo is based on FlowDroid~\cite{FlowDroid}, which provides a dataflow analysis~\cite{IFDS} framework for Android. In our implementation, we build a dummy main entry, and perform static code analysis based on our customized rules. For better graphical user interface (GUI) objects and user-driven callbacks analyses, we use GATOR\footnote{http://web.cse.ohio-state.edu/presto/software/gator/} which offers a more precise inter-component communication analysis. The above tools are based on Soot~\cite{Soot}, which is a framework for analyzing and transforming Java and Android applications.

\section{Evaluation}
\label{sec:evaluation}
In this section, we first introduce our experimental methodology, and then present the results of a small-scale real deployment and trace-driven simulations.

%\textbf{RQ1:} Can our system work with real-world applications, and what about the performance of code offloading on the cloudlet and the cloud?
%
%\textbf{RQ2:} What about the performance of extended functions, including application-specific data cache and object proxy mechanism?
%
%\textbf{RQ3:} If the scheduling strategy is effective when large number of requests arrive?

\begin{figure*}
\centering
\begin{subfigure}[tb]{0.48\textwidth}
\centering
\includegraphics[width=0.8\textwidth]{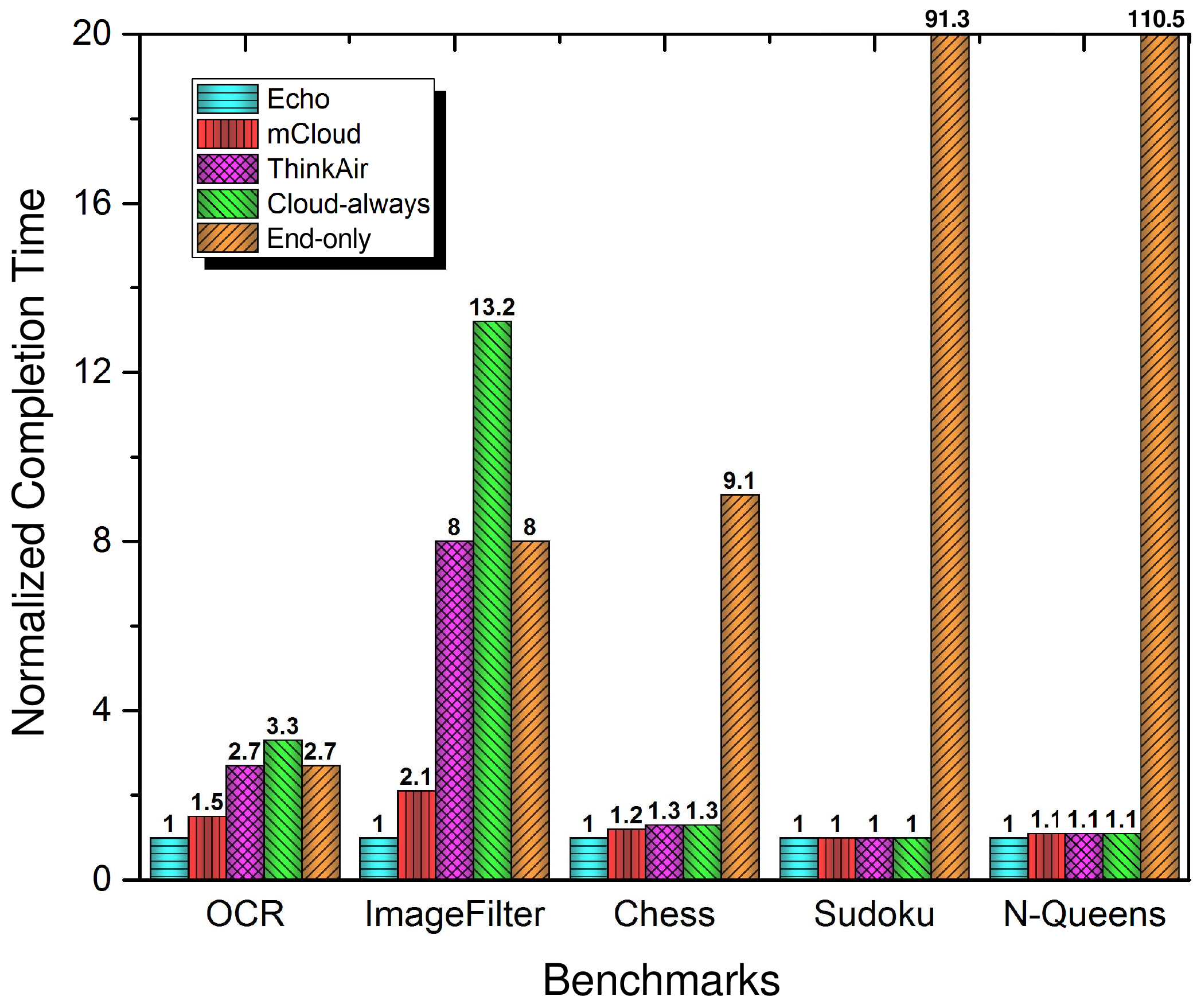}
\caption{Normalized task completion time}
\label{fig_benchmarks_time}
\end{subfigure}
\begin{subfigure}[tb]{0.48\textwidth}
\centering
\includegraphics[width=0.8\textwidth]{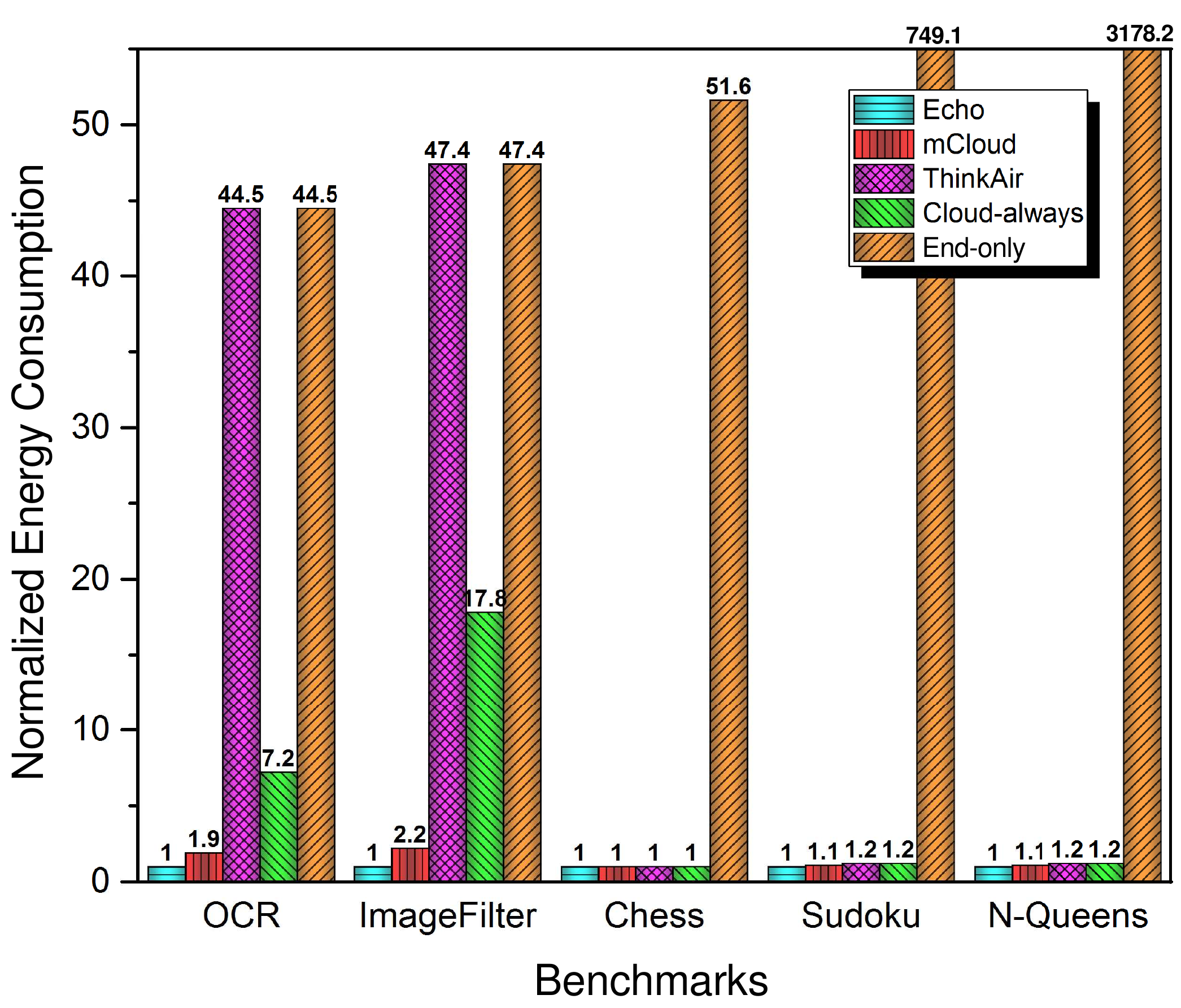}
\caption{Normalized energy consumption}
\label{fig_benchmarks_energy}
\end{subfigure}
\caption{Normalized task completion time and energy consumption}
\label{fig_benchmarks_timeandenergy}
\end{figure*}

\subsection{Methodology}
We build a small-scale testbed in our laboratory building to evaluate the performance of Echo. As shown in Fig.~\ref{fig:testbed}, we set up an edge node equipped with i7-4720HQ (2.60GHz with 4 cores of 8 threads) and 16GB RAM. The edge node is installed with Windows 8.1, and it launches Android-x86 VMs using VMWare Workstation. Android-x86 is a project that ports Android OS to x86 platforms. Each Android-x86 VM is assigned 2 virtual cores and 4GB memory. We organize 10 people to run mobile applications in the experiments, and each one has a Samsung Galaxy N7000 smartphones with a dual-core 1.4GHz CPU and Android OS 4.4. Smartphones connect to the edge node via WiFi network, which achieves a better offloading performance compared to 3G and 4G network~\cite{EmpiricalStudyOfEdgeApplications, LTEEnergy}. We rent VM instances with 2-core CPU and 4GB memory at AlibabaCloud\footnote{https://www.alibabacloud.com}. The network bandwidth of each VM instance at AlibabaCloud is 1Mbps.

%To achieve a better offloading performance, we use WiFi network, which outperforms 3G and 4G~\cite{EmpiricalStudyOfEdgeApplications, LTEEnergy} in terms of method execution time and energy consumption.

Mobile users run five open-source applications as benchmarks, including three interactive applications and two classic compute-intensive ones, which are summarized in Table~\ref{tab_benchmarks}. %Figure~\ref{fig_ocr} demonstrates the use of the OCR application.
We modify their source code by annotating time-consuming methods to be offloaded. For example, in OCR, we choose the method that performs the text extraction algorithm. We use PowerTutor~\cite{PowerTutor} to measure the energy consumption of smartphones.

%In simulations, we collect traces by running five applications with different user operations. For each annotated method, we measure its running time at different platforms as well as data uploading and downloading time. We compare Echo with two strategies --- End-only and Cloud-always, and the existing offloading system ThinkAir~\cite{ThinkAir}.
For comprehensive evaluations, we compare Echo with the following systems.
\begin{itemize}
\item End-only: all methods are executed on mobile devices.
\item Cloud-always: all annotated methods are offloaded to the cloud, and it is implemented based on ThinkAir.
\item ThinkAir~\cite{ThinkAir}: only cloud is available for code offloading. Each annotated method is offloaded to the cloud if offloading is faster than local execution. Otherwise, this method is executed by the mobile device.
\item mCloud~\cite{mCloud}: an edge assisted offloading system whose offloading decisions are made by mobile devices individually (different from Echo), and it offers a best-effort service at the edge without data transmission optimization. Echo and mCloud are both three-layer offloading systems, but Echo performs an edge-centric offloading decision with PC-SRTF.
\end{itemize}

\begin{table}[t]
\newcommand{\tabincell}[2]{\begin{tabular}{@{}#1@{}}#2\end{tabular}}
\newcommand{\PreserveBackslash}[1]{\let\temp=\\#1\let\\=\temp}
\newcolumntype{C}[1]{>{\PreserveBackslash\centering}p{#1}}
\newcolumntype{R}[1]{>{\PreserveBackslash\raggedleft}p{#1}}
\newcolumntype{L}[1]{>{\PreserveBackslash\raggedright}p{#1}}
\centering
\caption{Benchmarks}
\label{tab_benchmarks}
\begin{tabular} {|c|l|L{5cm}|}
\hline
 & \textbf{Benchmarks} & \textbf{Description} \\
\hline
\multirow{3}{*}{\begin{sideways}{\tabincell{cc}{\textbf{Interactive}}}\end{sideways}} & {Android-OCR} & Android optical character recognition, extracts the text from the image\\ \cline{2-3}
 & ImageFilter & Image filter for Android, applies filters to render images \\ \cline{2-3}
 & Chess & Chess game, enables man-machine playing and employs BFS to find the best moves  \\
\hline
\multirow{2}{*}{\begin{sideways}{\tabincell{cc}{\textbf{Compute}\\\textbf{Intensive}}}\end{sideways}}
 & Sudoku & A sudoku solver, uses a recursive descent backtracking algorithm to solve sudoku puzzles \\ \cline{2-3}
 & N-Queens & The classic N-Queens solver, uses a brute-force search algorithm to solve N-Queens puzzles \\
\hline
\end{tabular}
\end{table}

\begin{table*}[tb]
\centering
\caption{Task completion time and the energy consumption of CPU and WiFi}
\label{tab_time_energy_data}
\begin{tabular}{|l|l|l|l|l|l|l|l|l|l|l|}
\hline
\multicolumn{1}{|c}{\multirow{3}{*}{Benchmarks}} & \multicolumn{3}{|c}{Completion Time (s)} & \multicolumn{7}{|c|}{Energy Consumption (Joules)} \\ \cline{2-11}
\multicolumn{1}{|c}{} & \multicolumn{1}{|l}{\multirow{2}{*}{End-only}} & \multicolumn{1}{|l}{\multirow{2}{*}{Echo}} & \multicolumn{1}{|l}{\multirow{2}{*}{Cloud-always}} & \multicolumn{1}{|l|}{\multirow{2}{*}{End-only}} & \multicolumn{3}{c|}{Echo} & \multicolumn{3}{c|}{Cloud-always} \\ \cline{6-11}
\multicolumn{1}{|c}{} & \multicolumn{1}{|l}{} & \multicolumn{1}{|l}{} & \multicolumn{1}{|l}{} & \multicolumn{1}{|l|}{} & \multicolumn{1}{l}{CPU} & \multicolumn{1}{|l}{WiFi} & \multicolumn{1}{|l|}{Total} & \multicolumn{1}{l}{CPU} & \multicolumn{1}{|l}{WiFi} & \multicolumn{1}{|l|}{Total} \\ \hline
\multicolumn{1}{|l}{OCR} & \multicolumn{1}{|c}{15.14} & \multicolumn{1}{|l}{5.65} & \multicolumn{1}{|c}{18.77} & \multicolumn{1}{|c|}{11.58} & \multicolumn{1}{l}{0.03} & \multicolumn{1}{|l}{0.23} & \multicolumn{1}{|l|}{0.26} & \multicolumn{1}{l}{0.20} & \multicolumn{1}{|l}{1.68} & \multicolumn{1}{|l|}{1.88} \\ \hline
\multicolumn{1}{|l}{ImageFilter} & \multicolumn{1}{|c}{19.07} & \multicolumn{1}{|l}{2.37} & \multicolumn{1}{|c}{31.30} & \multicolumn{1}{|c|}{10.90} & \multicolumn{1}{l}{0.01} & \multicolumn{1}{|l}{0.22} & \multicolumn{1}{|l|}{0.23} & \multicolumn{1}{l}{0.52} & \multicolumn{1}{|l}{3.58} & \multicolumn{1}{|l|}{4.10} \\ \hline
\multicolumn{1}{|l}{Chess} & \multicolumn{1}{|c}{10.60} & \multicolumn{1}{|l}{1.17} & \multicolumn{1}{|c}{1.57} & \multicolumn{1}{|c|}{6.71} & \multicolumn{1}{l}{0.01} & \multicolumn{1}{|l}{0.12} & \multicolumn{1}{|l|}{0.13} & \multicolumn{1}{l}{0.01} & \multicolumn{1}{|l}{0.12} & \multicolumn{1}{|l|}{0.13} \\ \hline
\multicolumn{1}{|l}{Sudoku} & \multicolumn{1}{|c}{280.38} & \multicolumn{1}{|l}{3.07} & \multicolumn{1}{|c}{3.14} & \multicolumn{1}{|c|}{247.19} & \multicolumn{1}{l}{0.21} & \multicolumn{1}{|l}{0.12} & \multicolumn{1}{|l|}{0.33} & \multicolumn{1}{l}{0.28} & \multicolumn{1}{|l}{0.13} & \multicolumn{1}{|l|}{0.41} \\ \hline
\multicolumn{1}{|l}{N-Queens} & \multicolumn{1}{|c}{999.79} & \multicolumn{1}{|l}{9.05} & \multicolumn{1}{|c}{9.82} & \multicolumn{1}{|c|}{603.85} & \multicolumn{1}{l}{0.04} & \multicolumn{1}{|l}{0.15} & \multicolumn{1}{|l|}{0.19} & \multicolumn{1}{l}{0.07} & \multicolumn{1}{|l}{0.15} & \multicolumn{1}{|l|}{0.22} \\ \hline
\end{tabular}
\end{table*}

%\begin{figure}
%\centering
%\includegraphics[width=0.35\textwidth]{figs/ocr-image.jpg}
%\caption{The demo of the OCR application}
%\label{fig_ocr}
%\end{figure}

%\subsection{Experimental results}

%\begin{figure*}
%\centering
%\begin{subfigure}[tb]{0.45\textwidth}
%\includegraphics[width=0.8\textwidth]{figs/ExeTimeRatio.eps}
%\caption{Normalized task completion time}
%\label{fig_benchmarks_time}
%\end{subfigure}
%\begin{subfigure}[tb]{0.45\textwidth}
%\includegraphics[width=0.8\textwidth]{figs/EnergyRatio.eps}
%\caption{Normalized energy consumption of smartphone}
%\label{fig_benchmarks_energy}
%\end{subfigure}
%\caption{Normalized task completion time and smartphone energy consumption under different applications}
%\label{fig_benchmarks_timeandenergy}
%\end{figure*}

\subsection{Performance of Real Deployment}
\subsubsection{Overall performance}
We first study the task completion time by normalizing the results of all systems to the performance of Echo. For each application, the results are the average of 10 test runs (one per person) under the five system designs. As shown in Fig.~\ref{fig_benchmarks_timeandenergy}(a), Echo significantly outperforms other systems under interactive applications. Compared to ThinkAir, Echo can reduce the average completion time by about 63\% and 88\% under OCR and ImageFilter, respectively. Meanwhile, cloud-always has the worst performance under these two applications.
The offloaded methods in OCR and ImageFilter usually involve heavy data transmission but moderate computation. Therefore, Echo and mCloud show great advantages because of quick data uploading to edge. Echo can further outperform mCloud, thanks to our proposed data transmission optimization. Under compute-intensive applications, i.e., Sudoku and N-Queens, Echo achieves similar performance with cloud-always and ThinkAir, but they significantly outperform end-only. That is because both applications have little data for uploading and the benefits of edge proximity are not obvious. Moreover, since there is little resource contention at edge VMs in small-scale experiments, methods performing compute-intensive tasks have similar execution time at edge and cloud with similar VM configurations.

The normalized energy consumption of smartphone is shown in Fig.~\ref{fig_benchmarks_timeandenergy}(b). Under OCR and ImageFilter, Echo and mCloud save more energy than other systems because they run few methods on smartphones; meanwhile, the energy consumption of data transmission to the edge is more efficient than to the cloud. Moreover, Echo is more energy-efficient than mCloud due to less data transmission. Table~\ref{tab_time_energy_data} provides more details by showing the energy consumption of CPU and network interface. Under OCR and ImageFilter, Echo saves 86.3\% and 93.9\% WiFi energy, respectively, compared to cloud-always.

\begin{table}[tb]
\newcommand{\tabincell}[2]{\begin{tabular}{@{}#1@{}}#2\end{tabular}}
\newcommand{\PreserveBackslash}[1]{\let\temp=\\#1\let\\=\temp}
\newcolumntype{C}[1]{>{\PreserveBackslash\centering}p{#1}}
\newcolumntype{R}[1]{>{\PreserveBackslash\raggedleft}p{#1}}
\newcolumntype{L}[1]{>{\PreserveBackslash\raggedright}p{#1}}
\centering
\caption{Comparison of the transmission data, time and energy consumption with and without the lazy object transmission}
\label{tab_proxy_data}
\begin{tabular}{|l|l|l|}
\hline
& \tabincell{cc}{Without lazy \\ object transmission}   & \tabincell{cc}{With lazy \\ object transmission} \\ \hline
Transmission Data & 1951.0 (KB) & 262.9 (KB) \\ \hline
Transmission Time & 3.9 (S) & 0.5 (S) \\ \hline
Transmission Energy & 450 (mJ) & 116 (mJ) \\ \hline
\end{tabular}
\end{table}

\begin{figure}[tb]
\centering
\includegraphics[width=0.30\textwidth]{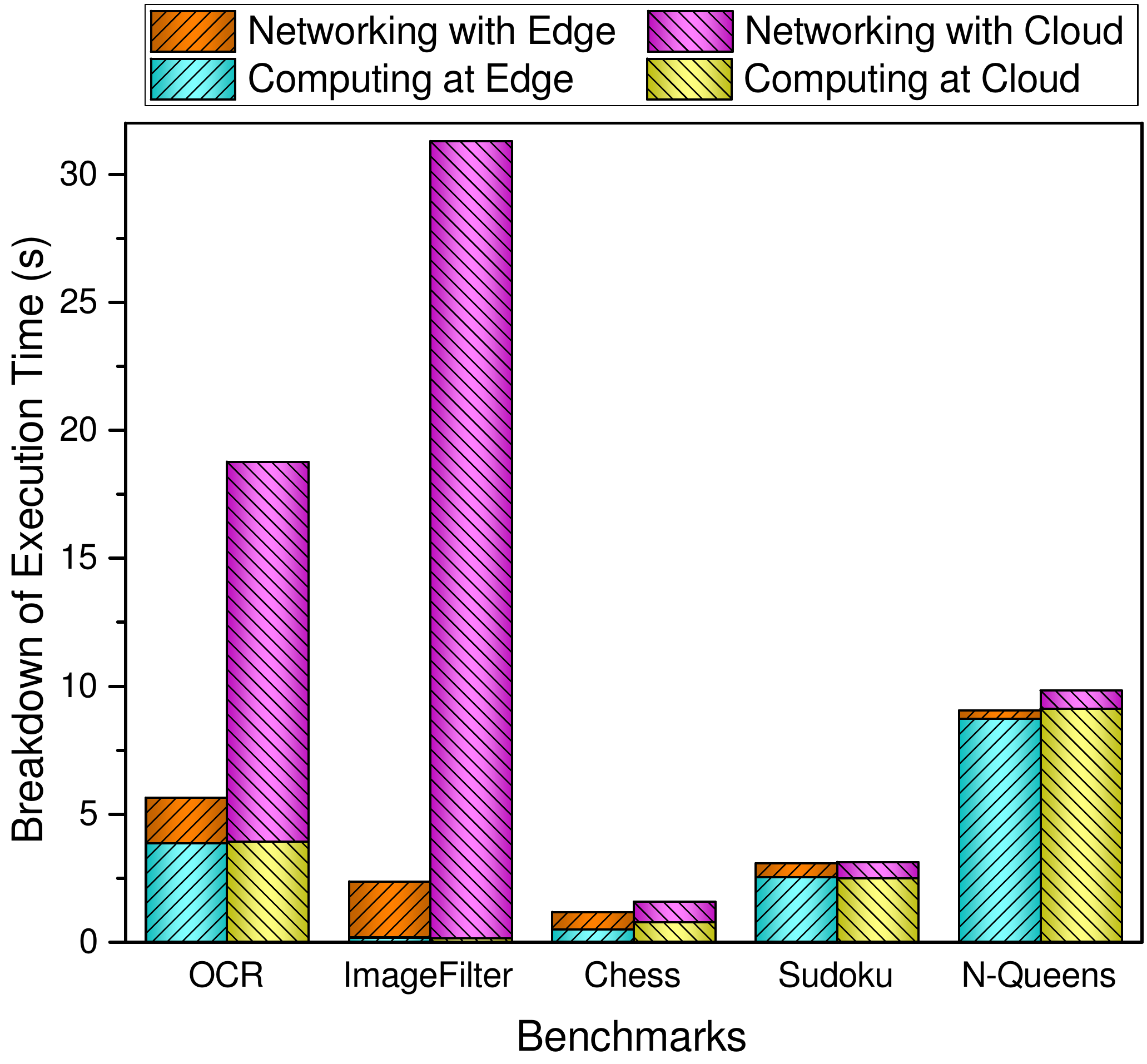}
\caption{The execution time breakdown of code offloading performing at edge and at cloud.}
\label{fig_time_breakdown}
\end{figure}

We further study the execution time breakdown of methods offloaded to the edge and cloud. As shown in Fig.~\ref{fig_time_breakdown}, the left bar and the right bar in each group show time breakdown between computation and network transmission at edge and at cloud respectively. Thanks to the proximity of edge node, smartphones can always quickly get the results by offloading methods to edge. Under interactive applications, a large portion of time is spent on data transmission. This phenomenon is more obvious when methods are offloaded to cloud. For example, 99.6\% time is spent on data transmission under ImageFilter. In contrast, computing occupies most of time in compute-intensive applications.

\begin{figure}[!tb]
\centering
\includegraphics[width=0.30\textwidth]{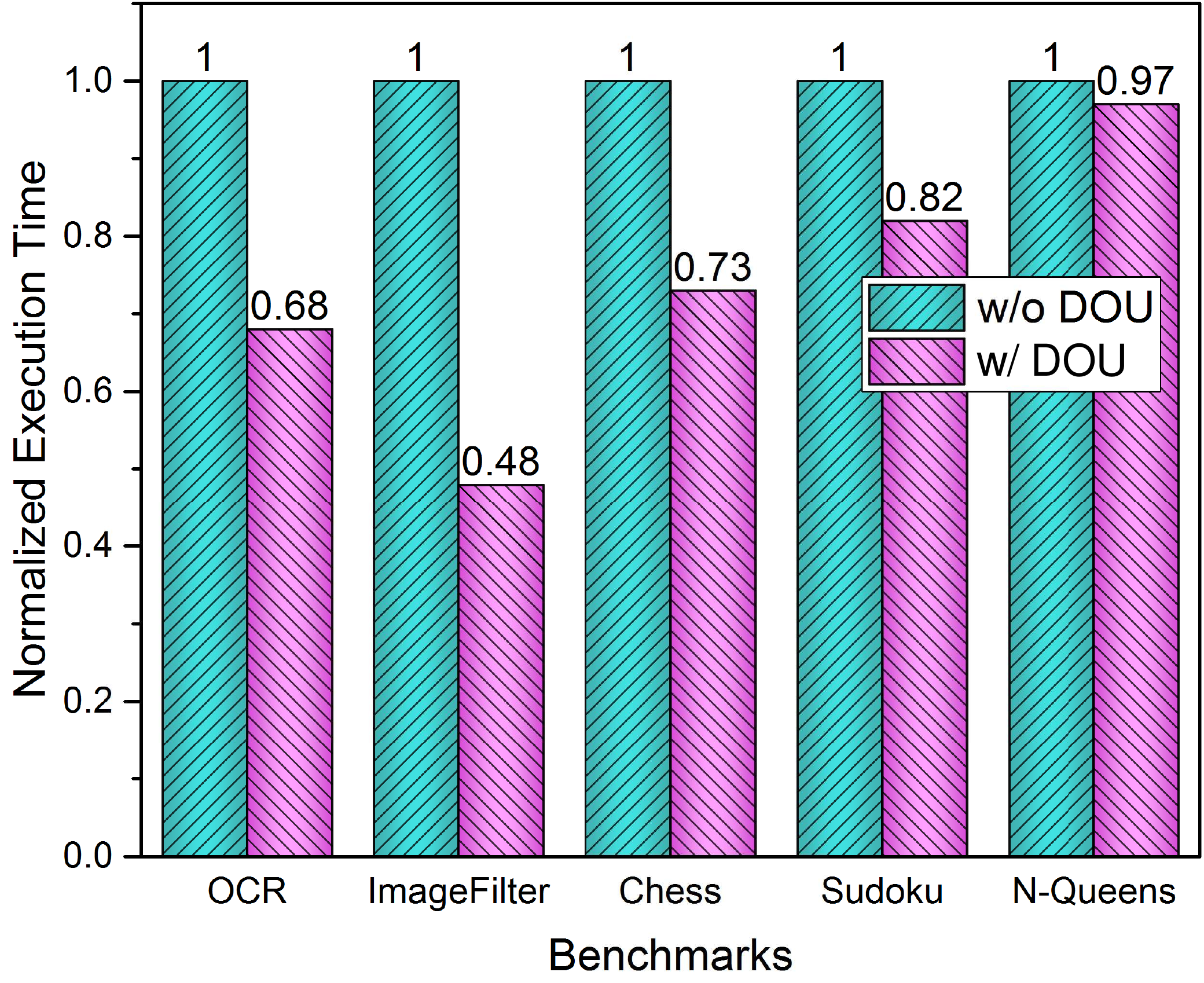}
\caption{Normalized execution time of Echo without differential object update (w/o DOU) and with differential object update (w/ DOU)}
\label{fig:cache:performance}
\end{figure}

\begin{figure}[!tb]
\centering
\includegraphics[width=0.30\textwidth]{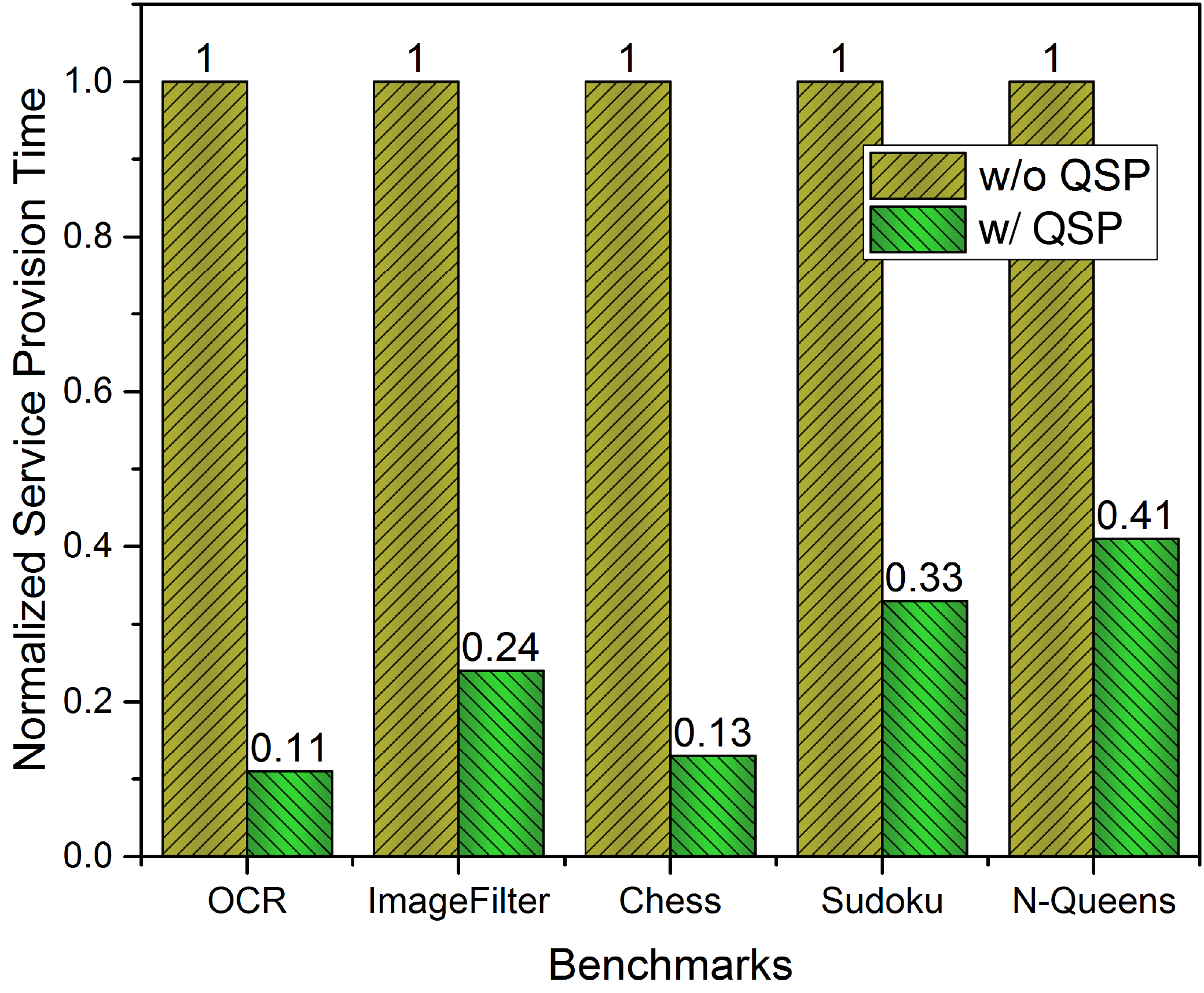}
\caption{Normalized service provision time of Echo without quick service provision (w/o QSP) and with quick service provision (w/ QSP)}
\label{fig:provision:performance}
\end{figure}

\begin{figure*}[tb]
    \centering
    \shortstack{
            \includegraphics[width=0.32\textwidth, height=0.3\linewidth]{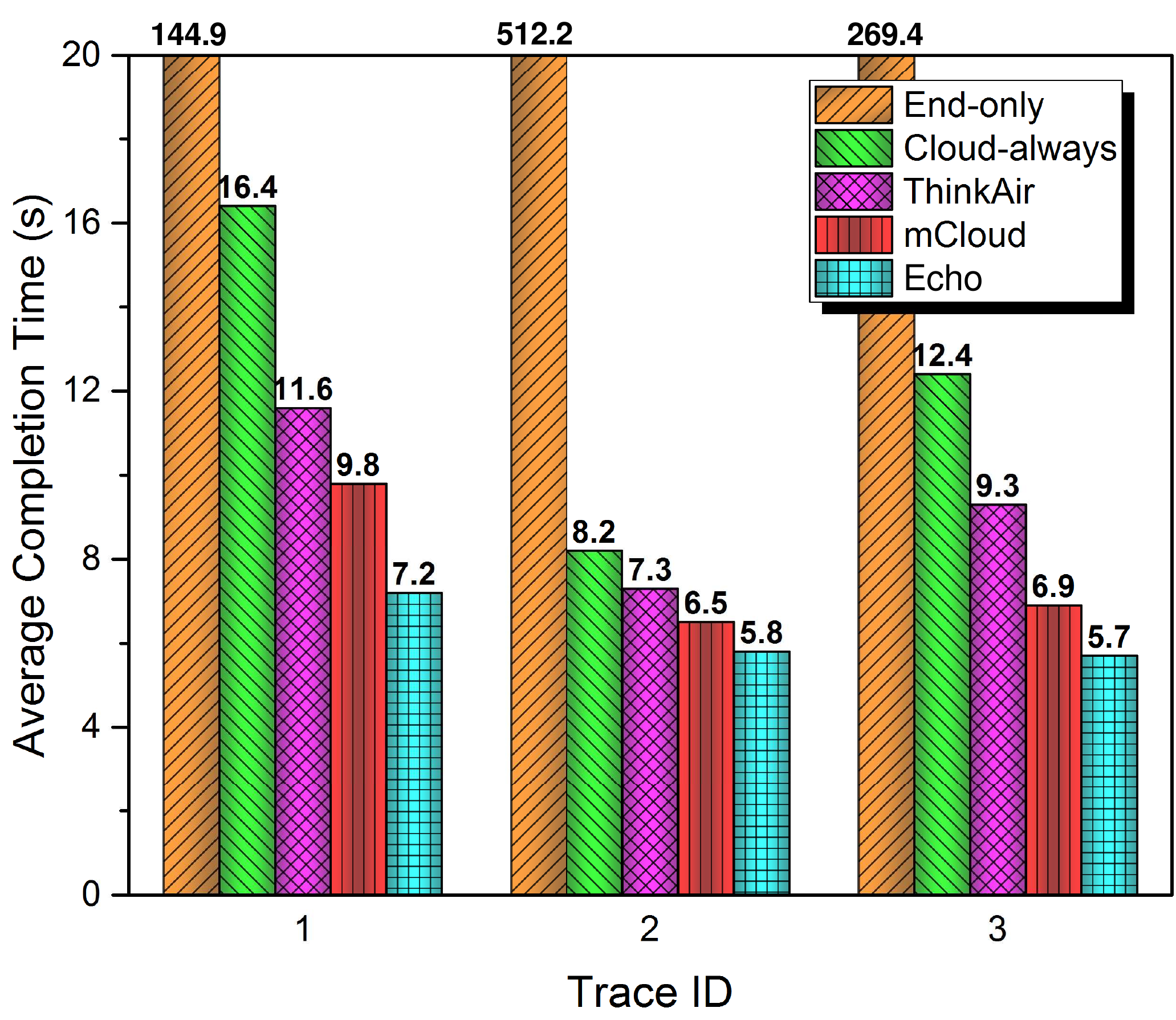}\\
            {(a) 1 VM}
    }\quad
    \shortstack{
            \includegraphics[width=0.32\textwidth, height=0.3\linewidth]{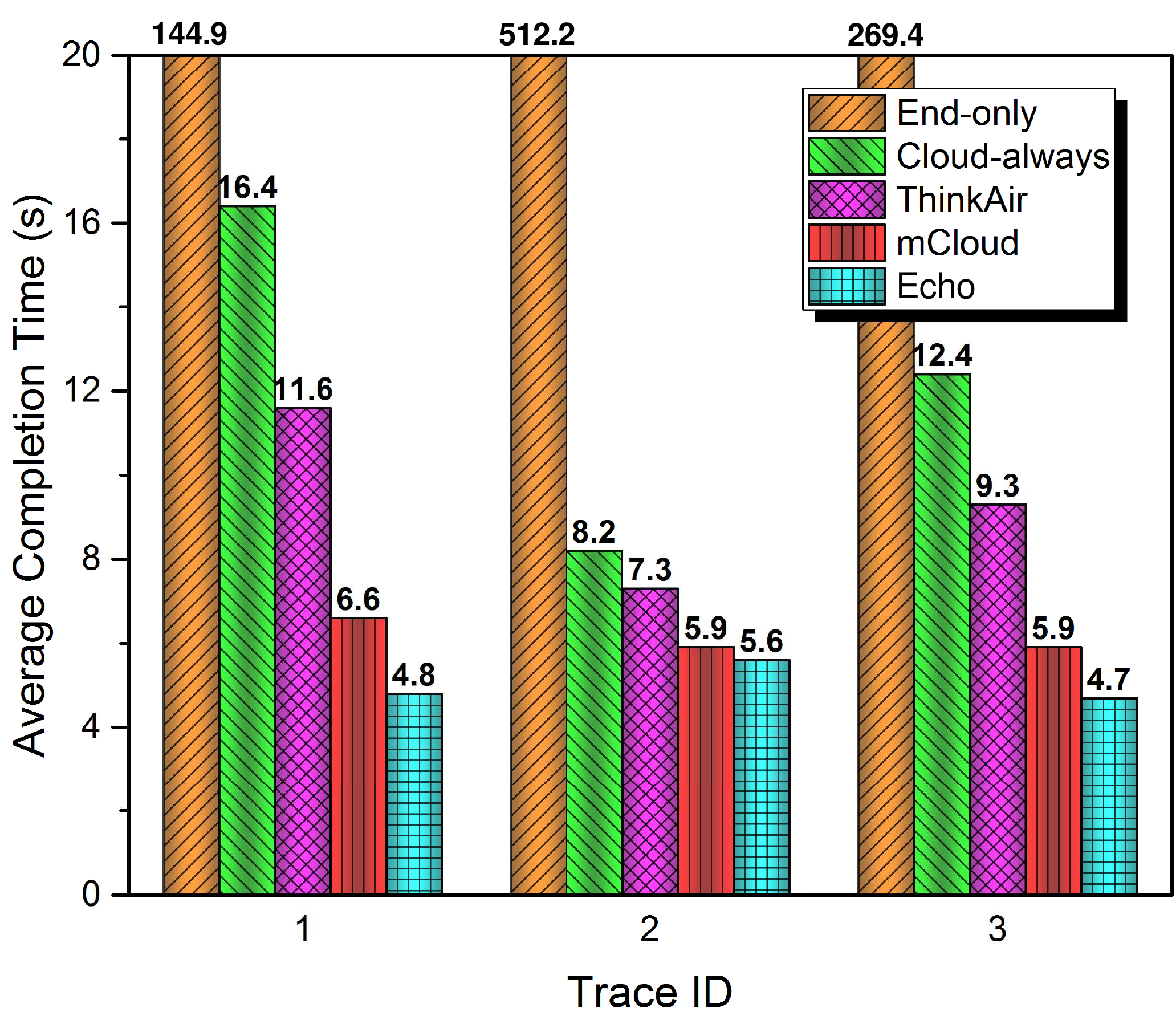}\\
            {(b) 4 VMs}
    }\quad
    \shortstack{
            \includegraphics[width=0.32\textwidth, height=0.3\linewidth]{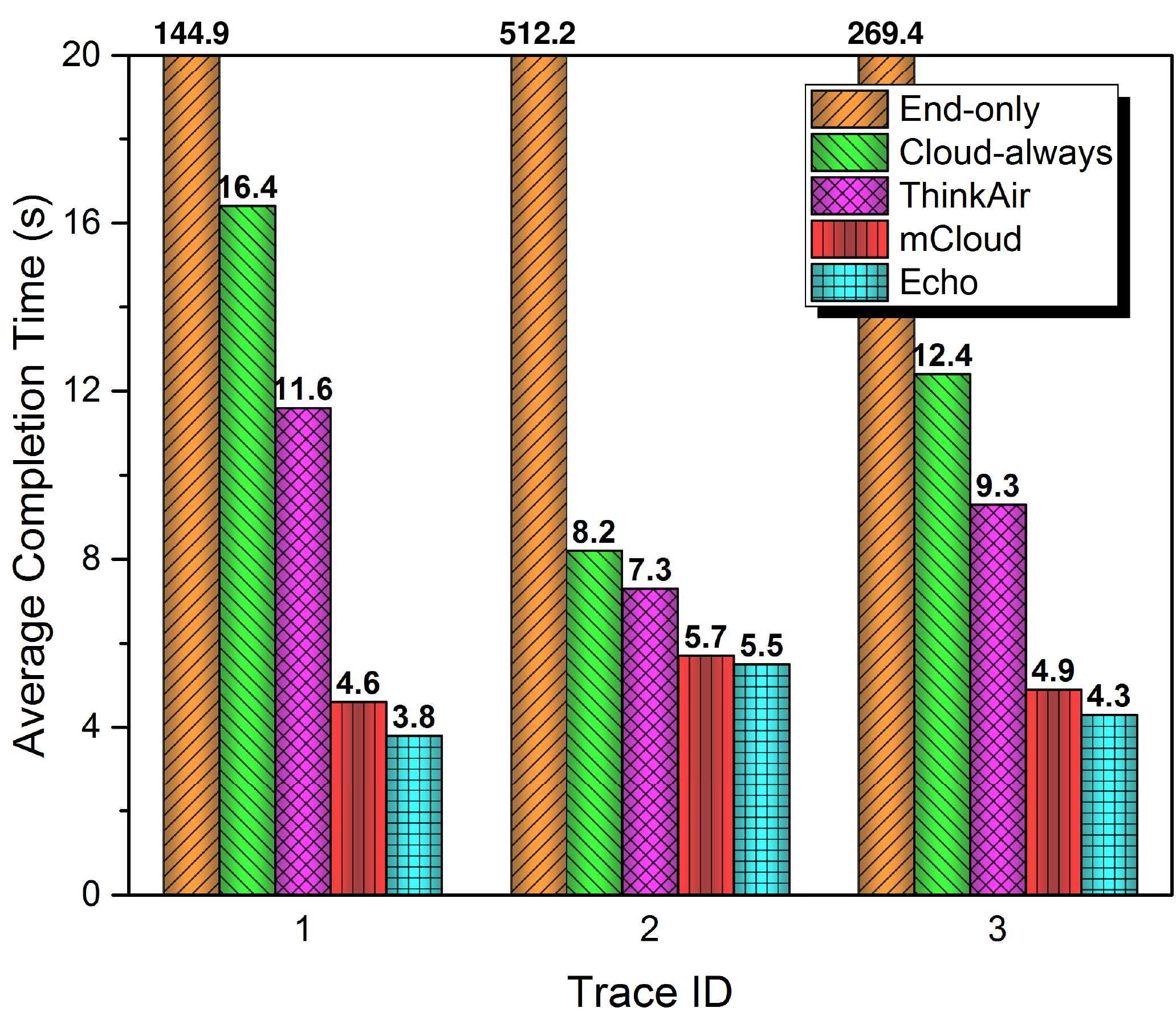}\\
            {(c) 8 VMs}
    }
    \caption{Average completion time when $\lambda=1$}
    \label{fig:Simulation1}
\end{figure*}
\begin{figure*}[tb]
    \centering
    \shortstack{
            \includegraphics[width=0.32\textwidth, height=0.3\linewidth]{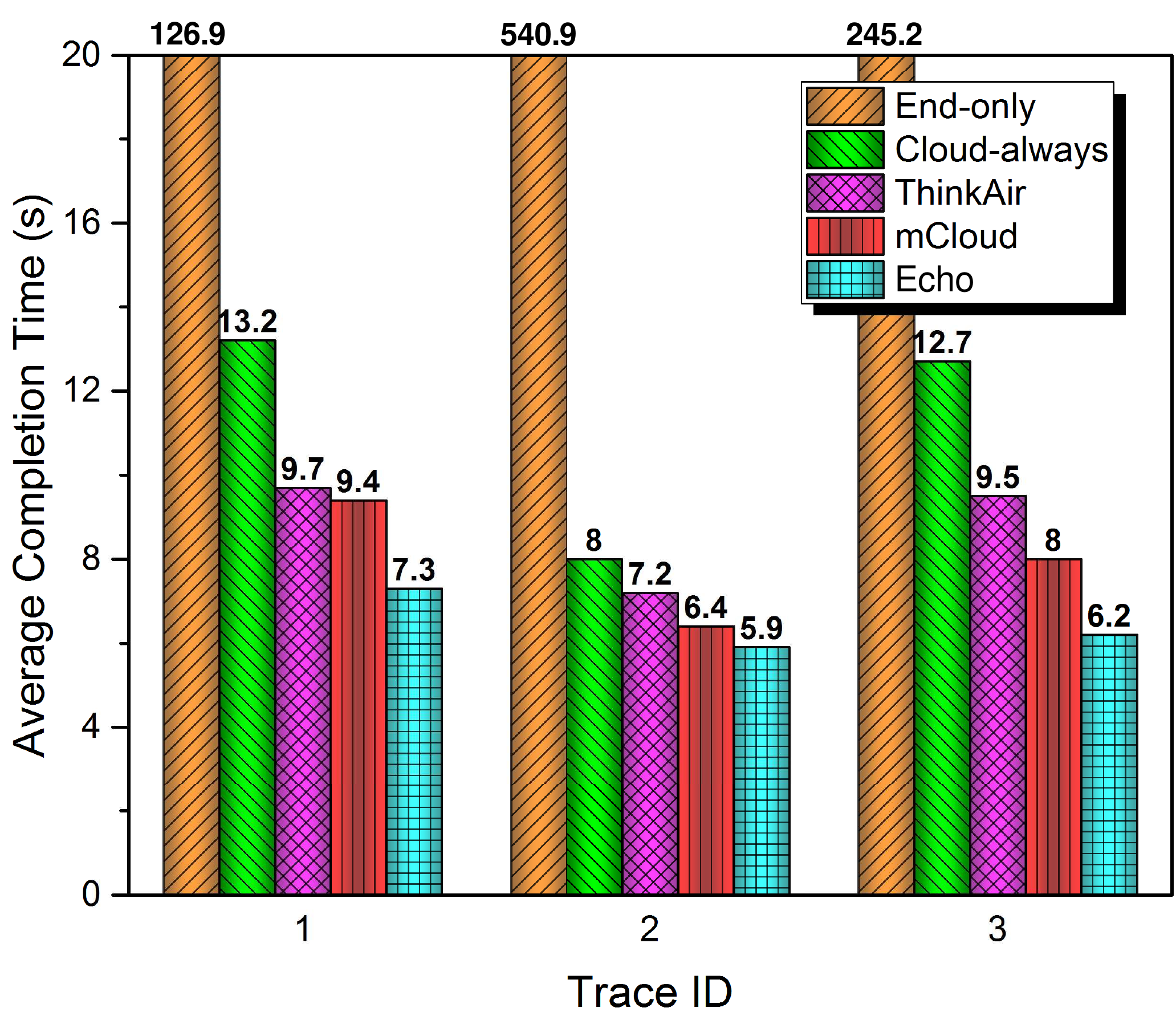}\\
            {(a) 1 VM}
    }\quad
    \shortstack{
            \includegraphics[width=0.32\textwidth, height=0.3\linewidth]{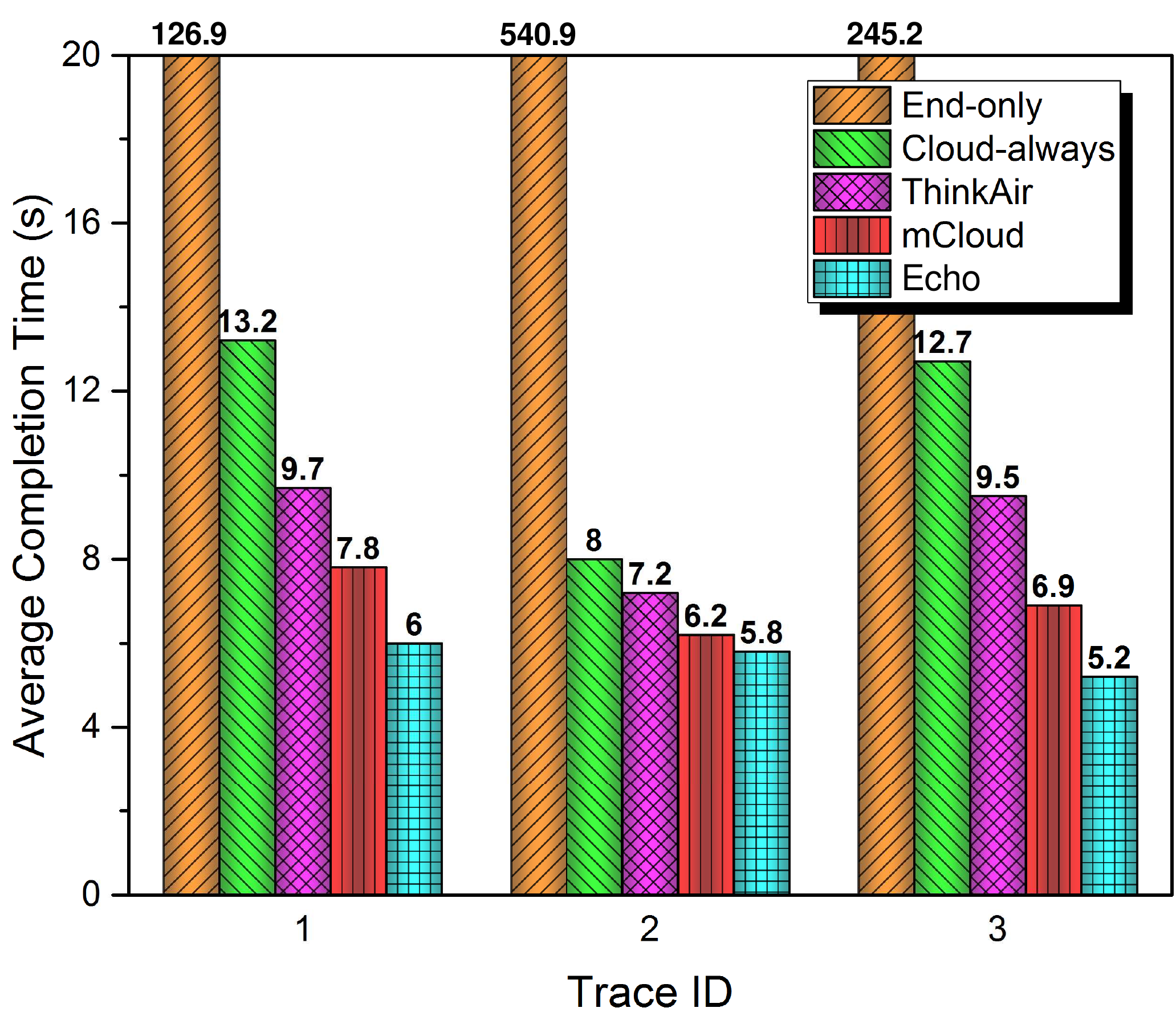}\\
            {(b) 4 VMs}
    }\quad
    \shortstack{
            \includegraphics[width=0.32\textwidth, height=0.3\linewidth]{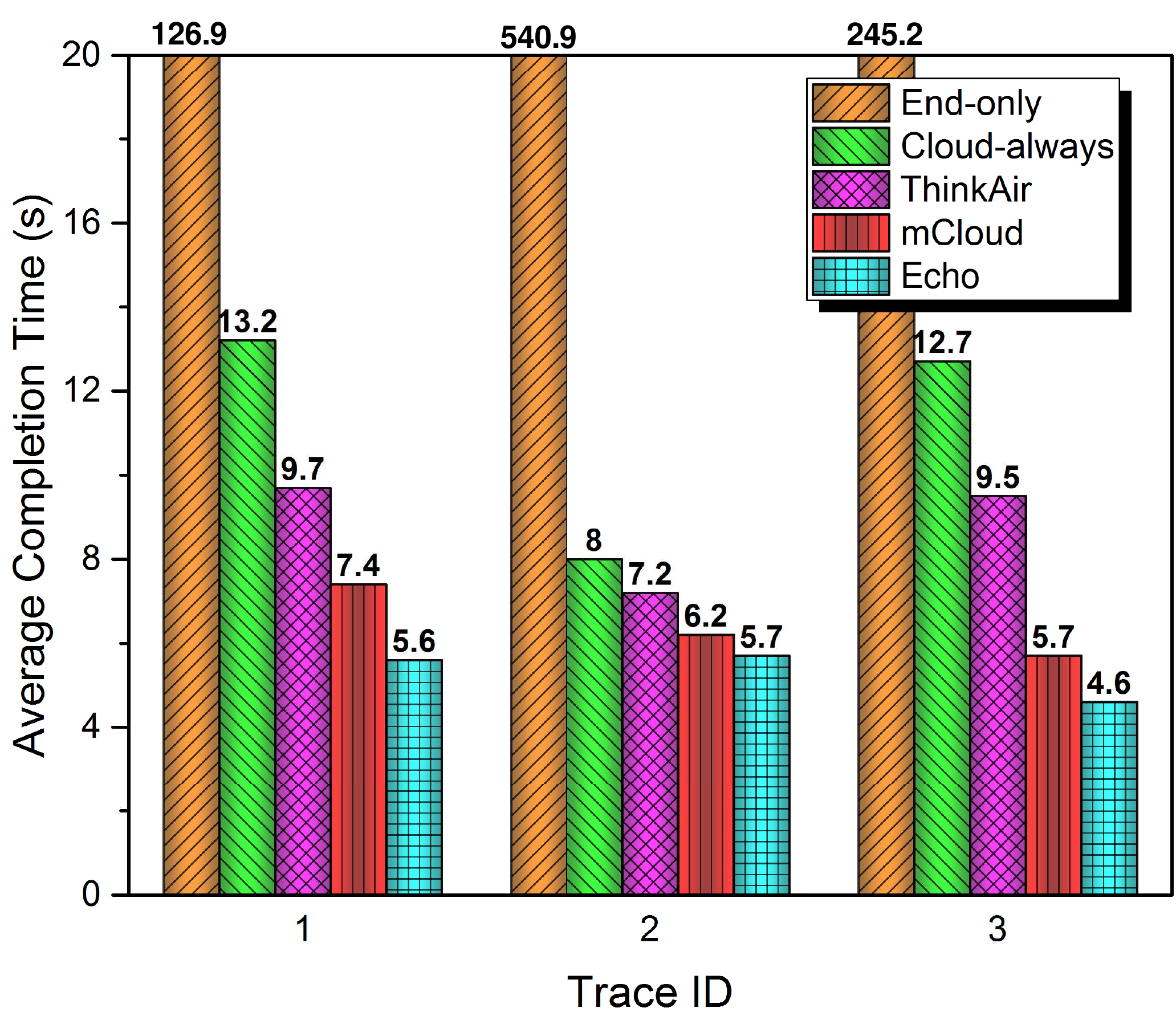}\\
            {(c) 8 VMs}
    }
    \caption{Average completion time when $\lambda=2$}
    \label{fig:Simulation2}
\end{figure*}

\subsubsection{Performance of lazy object transmission}

We use the Android-OCR application as a case study to evaluate the performance improvement by lazy object transmission (in Section~\ref{sec:lazy:transmission}). In the implementation of Android-OCR, it first searches the mobile storage to load the image set, which contains images to be selected for text extraction. When the method responsible for text extraction is offloaded, the original image set is transmitted to the server. However, the images selected by users can only be determined at runtime. In this case, the lazy object transmission can effectively avoid transmitting images that are not selected by users so as to remove unnecessary data transmission. Table~\ref{tab_proxy_data} compares the performance of Android-OCR with and without the lazy object transmission. In this example, the original image set includes 6 pre-installed images. If the user chooses only one image for text extraction, we can significantly reduce data transmission time and energy consumption.

\subsubsection{Performance of differential object update}
In Echo, we use the differential object update (DOU, in Section~\ref{sec:differential:update}) to reduce the amount of data transmission during code offloading. In DOU, object data are cached at both mobile and remote (edge or cloud), and only the differential part of object needs to be transmitted when an object updates. We measure the execution time of offloaded methods with and without DOU, respectively, and show the normalized results in Fig.~\ref{fig:cache:performance}. We observe that method execution with DOU can significantly reduce method execution time, especially in interactive applications, which perform more data exchange with remote servers.
%The results are consistent with those in Fig.~\ref{fig_time_breakdown}.

\subsubsection{Performance of quick service provision}
We study the performance of quick service provision (QSP, details in Section~\ref{sec:service:provision}) by measuring the service provision time, which is defined as a period between receiving an offloading request and being ready to run the task. Specifically, it includes time of VM initialization and user execution environment setup. As shown in Fig.~\ref{fig:provision:performance}, QSP can significantly reduce service provision time due to quick VM synthesis and application package pre-installation.

\subsection{Trace-driven Simulations}
We develop a trace-driven simulator to evaluate Echo at a larger scale. First, we collect traces from mobile users in real deployment, who run five applications with different operations. For each annotated offloading method, we measure its running time at different platforms as well as data uploading and downloading time. Then, we choose three sets of traces by mixing five applications with different proportions. Each set contains 100 offloading requests. The first set of traces contains 80\% methods from interactive applications (listed in Table~\ref{tab_benchmarks}) and 20\% methods from compute-intensive ones. In the second set of traces, we reverse the proportion by including 20\% interactive applications and 80\% compute-intensive ones. The proportion is 50\% for both in the third set. We adopt a widely-adopted service model that the interval time between two requests is an exponential distribution with parameter $\lambda$, so that the number of requests in unit time is a normal distribution. In our simulations, there are 100 mobile users and the number of VMs at the edge node can be changed from 1 to 8.

We first set $\lambda=1$ and show the results in Fig. \ref{fig:Simulation1}. It is obviously that Echo outperforms other systems in all cases, and their performance gap increases as more VMs are available at the edge. However, we observe that the performance gap between Echo and mCloud becomes smaller when the number of VMs grows. That is because more VMs mitigate resource contention at the edge. We then increase resource contention by setting $\lambda=2$, i.e., the mean of time interval between two requests is 0.5 second. As shown in Fig. \ref{fig:Simulation2}, Echo brings more benefits. For example, when there are 8 VMs, Echo reduces the average completion time by 24.3\% under the first set of traces compared with mCloud, while the reduction is only 17.4\% when $\lambda=1$.

\section{Conclusion}
\label{sec:conclusion}

In this paper, we propose Echo, an edge-centric code offloading system over mobile devices, edge and cloud with QoS guarantee. It has a centralized decision engine that collects offloading requests from mobile devices and decides which methods should be offloaded to edge or cloud. To reduce average task completion time, we propose a heuristic algorithm called PC-SRTF for task scheduling at the edge. Echo also optimizes network transmission by lazy object transmission and differential object update. Through a small-scale real deployment and trace-driven simulations, we show that Echo can significantly outperform existing code offloading systems.


%\section*{Acknowledgment}




%\bibliographystyle{IEEEtran}
%\begin{thebibliography}{1}
%\end{thebibliography}

\bibliographystyle{IEEEtran}
\bibliography{sdo}

\end{document}